# Anomalous-Hall Néel textures in altermagnetic materials

Rui-Chun Xiao,[1,2,*] Hui Li,[1,2] Hui Han,[1,2] Wei Gan,[1] Mengmeng Yang,[1,2] Ding-Fu Shao,[3] Shu-Hui Zhang,[4] Yang Gao,[5] Mingliang Tian,[6,7,*] Jianhui Zhou[6,*]

[1] *Institute of Physical Science and Information Technology, Anhui University, Hefei 230601, China*
[2]*Anhui Provincial Key Laboratory of Magnetic Functional Materials and Devices, School of Materials Science and Engineering, Anhui University, Hefei 230601, China*
[3]*Key Laboratory of Materials Physics, Institute of Solid State Physics, Chinese Academy of Sciences, 230031 Hefei, China*
[4]*College of Mathematics and Physics, Beijing University of Chemical Technology, Beijing 100029, China*
[5]*Department of Physics, University of Science and Technology of China, Hefei 230026, China.*
[6]*Anhui Key Laboratory of Low-Energy Quantum Materials and Devices, High Magnetic Field Laboratory, HFIPS, Chinese Academy of Sciences, Hefei, Anhui 230031, China.*
[7]*School of Physics and Optoelectronics Engineering, Anhui University, Hefei 230601, China*
**Correspondence author:* xiaoruichun@ahu.edu.cn (R.-C. Xiao), mltian@ahu.edu.cn (M.L. Tian), jhzhou@hmfl.ac.cn (J.H. Zhou)

Recently, the altermagnets, a new kind of collinear antiferromagnet with nearly zero net magnetization and momentum-dependent spin-splitting of bands, have sparked great interest. Despite simple magnetic structures, these altermagnets exhibit intriguing and intricate dependence of anomalous Hall effect (AHE) on the Néel vector, in contrast to the conventional perpendicular configuration of Hall current with magnetization in ferromagnets. However, the fundamental relationship between the AHE and the Néel vector remains largely elusive. Here, we reveal all the unconventional anomalous Hall textures in the Néel vector space, dubbed anomalous-Hall Néel textures (AHNTs) for altermagnets. Specifically, we identify 10 types across four categories of AHNTs for all altermagnets. Notably, we find that AHNTs resemble the known spin textures in momentum space, and further reveal their symmetry origin. Meanwhile, we examine our key discoveries in prototypical altermagnets. Our work offers a thorough understanding of AHE in altermagnets and a complete and pictorial classification of altermagnets based on the geometry of response functions.

## 1 Introduction

The anomalous Hall effect (AHE) [1] and its optical counterpart magneto-optical effect (Kerr and Faraday effect) [2] are the salient manifestations of time-reversal symmetry breaking in magnetically ordered materials. The Hall current $\mathbf{j}_H = \boldsymbol{\sigma}_H \times \mathbf{E}$ is determined by anomalous Hall vector, $\boldsymbol{\sigma}_H = (\sigma_{Hx}, \sigma_{Hy}, \sigma_{Hz})$, and $\sigma_{Hi} = \varepsilon_{ijk}\sigma_{jk}^{A}$ $(i, j, k \in (x, y, z))$ is the anomalous Hall conductivity which is intimately linked to the Berry curvature of electrons in momentum space [3,4]. In ferromagnets and ferrimagnets, $\boldsymbol{\sigma}_H$ is usually believed to be parallel to the magnetization $\mathbf{M}$, that is, the radial structure ($\boldsymbol{\sigma}_H \parallel \mathbf{M}$) in the magnetization space (such as Fig. 1(a)), which leads to the conventional perpendicular configuration $\mathbf{j}_H \perp \mathbf{M}$. This long-time belief has been revisited by the recent progress of in-plane AHE [5-9]. Notably, AHE has been observed in nearly compensated noncolinear antiferromagnetic (AFM) materials [4,10-12]. Meanwhile, unconventional spin currents in [13,14] and field-free magnetization switching [15] in Kagome nonlinear AFM (such as $Mn_3Sn$ and $Mn_3Ge$) have been reported. Moreover, in these systems, the sum of the radial components $\boldsymbol{\sigma}_H \parallel \mathbf{M}$ in the different spin sublattices is negligible; however, the non-radial components may exist and exhibit complicated behaviors.

Altermagnets [16-23], classified within the realm of collinear AFM materials, feature vanishing net magnetization constrained by symmetry, but exhibit momentum-dependent spin-split band structures akin to ferromagnets [24-29]. Remarkably, AHE [30-40] and the magneto-optical effects [41,42] have been observed in these materials, inspiring intensive investigations [43-49]. Despite the relatively simple magnetic structures in altermagnetic materials, the relationship between $\boldsymbol{\sigma}_H$ and the Néel vector $\mathbf{n}$ (the unit vector indicating the magnetic direction difference between two spin sublattices) unexpectedly exhibits rich and some puzzling features [36,41,50,51], unlike the conventional radial structure ($\boldsymbol{\sigma}_H \parallel \mathbf{M}$) in ferromagnets. However, a proper theory to build the fundamental relation between $\boldsymbol{\sigma}_H$ and the Néel vector $\mathbf{n}$ for all altermagnets remains elusive.

In this work, we treat the Néel vector as an extrinsic parameter to investigate anomalous Hall configurations in the Néel vector space $\boldsymbol{\sigma}_H(\mathbf{n})$ for altermagnets, which can be dubbed as anomalous-Hall Néel textures (AHNTs). We further establish a complete classification of these AHNTs, and find AHNTs exhibit unusual relationships in analogy to the spin textures [52-54] in momentum space (Fig. 1), such as Dresselhaus-like, Rashba-like, persistent, and pure cubic textures. Furthermore, we classify the altermagnetic materials according to the types of AHNTs and further verify our main results in several typical altermagnets.

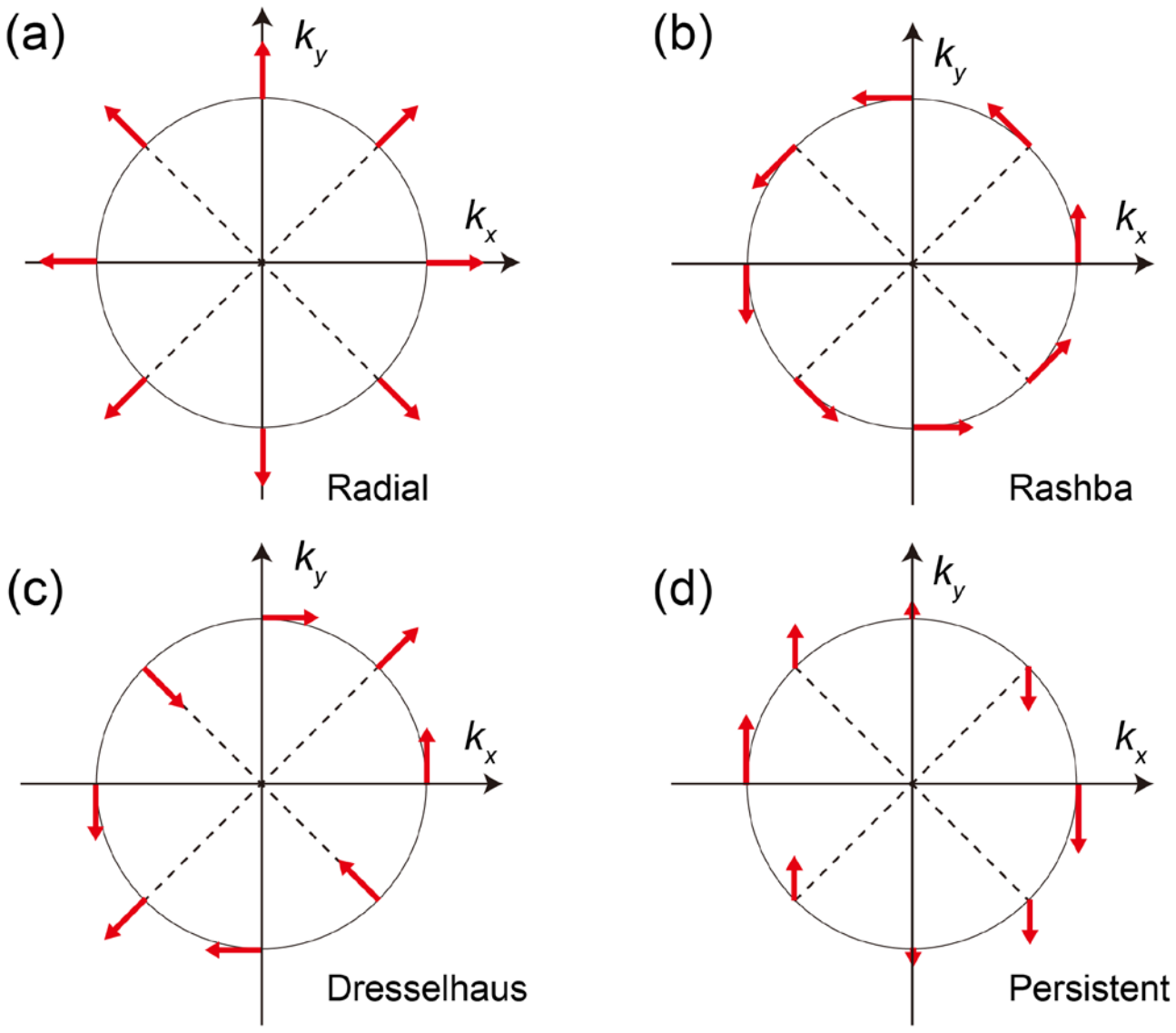

**Fig. 1.** Spin textures in the momentum space for nonmagnetic materials without inversion symmetry. (a) Radial $\beta_D(\sigma_x k_x + \sigma_y k_y)$, (b) Rashba $\lambda_R(k_x\sigma_y - k_y\sigma_x)$, (c) Dresselhaus $\lambda_D(k_x\sigma_y + k_y\sigma_x)$, (d) persistent $\lambda_R(k_x\sigma_y - k_y\sigma_x) + \lambda_D(k_x\sigma_y + k_y\sigma_x) = 2\lambda_R k_x\sigma_y$ with $\lambda_R = \lambda_D$ here. $\sigma_x$, $\sigma_y$, and $\sigma_z$ are the Pauli matrices.

## 2 Theoretical model

Since altermagnetic materials are all in symmetry-compensated collinear AFM states with nearly zero net spin magnetization, it is a natural way to analyze the AHE in terms of the Néel vector $\mathbf{n}$ [8,50,51], where $\mathbf{n}$ is a unit vector ($|\mathbf{n}| = 1$). To determine the profiles of $\boldsymbol{\sigma}_H(\mathbf{n})$, we treat $\mathbf{n}$ as an extrinsic parameter [55] rather than an intrinsic structural parameter in traditional magnetic group method (see Supplementary Note 2) [56], and we named it as "extrinsic parameter" method. This approach enables us to analyze the $\boldsymbol{\sigma}_H(\mathbf{n})$ relation in the framework of the nonmagnetic space group. The space group operation $\{R|\mathbf{t}\}$ acting on $\mathbf{n}$ results in the change from $\boldsymbol{\sigma}_H(\mathbf{n})$ to $\boldsymbol{\sigma}_H(\{R|\mathbf{t}\}\mathbf{n})$, and it is equivalent to apply the point symmetry operation $R$ directly on $\boldsymbol{\sigma}_H$, i.e.,

$$\boldsymbol{\sigma}_H(\{R|\mathbf{t}\}\mathbf{n}) = R\boldsymbol{\sigma}_H, \tag{1}$$

where the translation operation $\mathbf{t}$ is omitted on the right side, as it does not impact on macroscopic physical properties. Using Eq. (1), we can show that the antiferromagnets, in which two spin sublattices are connected either by inversion $P$ or pure translation $\mathbf{t}_c/2$, do not permit the existence of AHE, because these symmetries only reverse $\mathbf{n}$ not $\boldsymbol{\sigma}_H$ (Fig. 2(a) and (b)). Coincidentally, the breaking of these two symmetries [18,19,21,22,57] leads to the bands splitting without the spin-orbit coupling (SOC) effect, forming altermagnets. Notably, the spins are compensated, meaning the AHE in altermagnets is attributed to intrinsic orbital angular momenta [30].

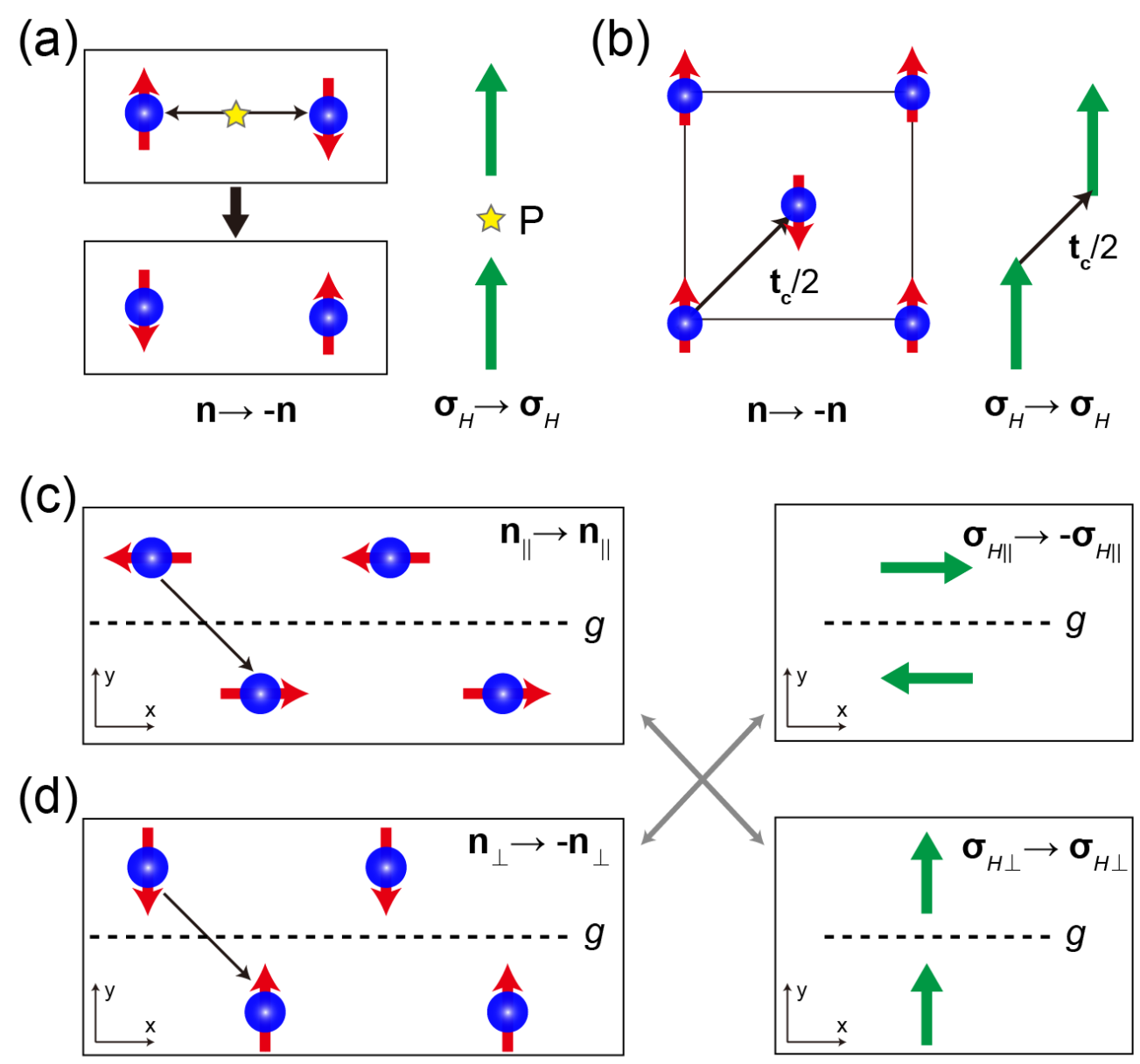

**Fig. 2.** Transformations of Néel vector $\mathbf{n}$ and anomalous Hall vector $\boldsymbol{\sigma}_H$ under the (a) inversion symmetry $P$, (b) $\mathbf{t}_c/2$ translation, and (c, d) glide mirror $g$ operations. The thin arrows between atoms denote the transformations of atom positions under the symmetry operations. The gray cross arrows between (c) and (d) denote the reasonable $\boldsymbol{\sigma}_H - \mathbf{n}$ relationships in altermagnets.

Under mirror or glide mirror symmetry, the relationship between $\boldsymbol{\sigma}_H$ and $\mathbf{n}$ turns out to be non-trivial. As shown in Fig. 2(c), when $\mathbf{n}$ is parallel to the glide mirror, the exchange of magnetic atoms with opposite spins preserves the Néel vector $\mathbf{n}_\parallel$ (see Supplementary Note 1). However, the glide mirror reverses the sign of $\boldsymbol{\sigma}_{H\parallel}$. Conversely, when the Néel vector is perpendicular to the glide mirror, the glide mirror operation reverses $\mathbf{n}_\perp$, while leaving the vertical component $\boldsymbol{\sigma}_{H\perp}$ unchanged as it is an axial vector, which is shown in **Fig.** 2(d). To ensure Eq. (1) holds, $\boldsymbol{\sigma}_H$ must be perpendicular to $\mathbf{n}$ in the cases of Fig. 2(c, d). Compared to general relationship $\boldsymbol{\sigma}_H \parallel \mathbf{m}$ in ferromagnets, $\boldsymbol{\sigma}_H \perp \mathbf{n}$ relation can emerge in altermagnets [36,41,50,51].

If the directions of the magnetic moments change continuously while remaining strictly anti-parallel between the two spin sublattices, the Néel vector $\mathbf{n}$ traces out a surface resembling a Bloch sphere $|\mathbf{n}| = 1$. We can regard $\boldsymbol{\sigma}_H - \mathbf{n}$ relation as the inner structure of $\boldsymbol{\sigma}_H$ in the Néel vector space. In Fig. 2 (c, d), we focus on two specific points in the Néel vector space: **n**=(1, 0, 0) and **n**=(0, 1, 0). Additional symmetries further lead to $\boldsymbol{\sigma}_H(\mathbf{n})$ exhibit unconventional configurations like spin textures in the momentum space. In analogy to the spin texture [52-54], we thus refer to the $\boldsymbol{\sigma}_H - \mathbf{n}$ relationship as the AHNT.

To gain analytical results of $\boldsymbol{\sigma}_H(\mathbf{n})$, we expand $\boldsymbol{\sigma}_H$ as a Taylor polynomial with respect to **n**:

$$\boldsymbol{\sigma}_H(\mathbf{n}) = \mathbf{T}^{(2)} \cdot \mathbf{n} + \mathbf{T}^{(4)} : \mathbf{nnn} + \cdots \qquad (2)$$

The Taylor expansion is guaranteed because the SOC interaction acts as a perturbation in AHE [8]. The Onsager reciprocal relation allows only the odd terms in Eq. (2), due to its oddness under time-reversal operation. $\mathbf{T}^{(2)}$and $\mathbf{T}^{(4)}$ are the two-rank and four-rank tensors, respectively. Alternatively, $\boldsymbol{\sigma}_H$ can be expressed in terms of components:

$$\sigma_{Hi} = \sum_{j=1}^{3} T_{ij}^{(2)} n_j + \sum_{j,k,l=1}^{3} T_{ijkl}^{(4)} n_j n_k n_l + \cdots \qquad (3)$$

The transformation of $\boldsymbol{\sigma}_H$ under the space group operation $\{R|\mathbf{t}\}$ in Eq. (1) becomes:

$$\{R|\mathbf{t}\}: \boldsymbol{\sigma}_H \to \det(R)\, D(R) \boldsymbol{\sigma}_H, \qquad (4)$$

where $D(R)$ is the matrix of symmetry operation $R$, and det($R$) is the determinant of $D(R)$. The transformation of the Néel vector **n** under $\{R|\mathbf{t}\}$ is given by:

$$\{R|\mathbf{t}\}: \mathbf{n} \to \pm \det|R| D(R) \mathbf{n}, \qquad (5)$$

where the + and - signs indicate whether the two magnetic sublattices with opposite spins are exchanged or not, respectively. Our method properly deals with the transformation of the Néel vector under the symmetry operations (see Supplementary Note 2). Although the time-reversal symmetry operation is not explicitly involved in this method, the ±1 symbol in Eq. (5) essentially reflects the switch of the magnetic moment under time-reversal operation.

With the aid of Eqs. (1)-(5), we can obtain the symmetry-adapted tensors $\mathbf{T}^{(2)}$ and $\mathbf{T}^{(4)}$, and identify the invariant polynomials of **n** in Eq. (3). Meanwhile, we have developed an automatically computational code to implement our method (see Supplementary Note 2).

## 3. Results and discussions

### *3.1 Classification of AHNTs for altermagnets.*

With the above extrinsic parameter method, we can conduct a comprehensive study on the $\boldsymbol{\sigma}_H(\mathbf{n})$ relations and corresponding AHNTs in the altermagnets. Since the tensors $\mathbf{T}^{(2)}$and $\mathbf{T}^{(4)}$ in Eq. (2) resembles the general axial tensors, it suffices to utilize the nonmagnetic group, rather than the magnetic group, to investigate the symmetry constraints imposed on $\mathbf{T}^{(2)}$ and $\mathbf{T}^{(4)}$ in the context of the "extrinsic parameter" method. We found 10 kinds of AHNTs in four categories for altermagnets, based on ten non-centrosymmetric and non-chiral point groups (see Supplementary Note 4). We can see that the Rashba-like, Dresselhaus-like and pure cubic AHNTs can exist in altermagnetic materials, while the radical texture which is common in ferromagnetic is absent.

As summarized in Table I, we find that the altermagnetic materials do not support the radial AHNTs as the ferromagnets due to the traceless characteristic of $\mathbf{T}^{(2)}$. If the symmetry composed on $\boldsymbol{\sigma}_H$ is equivalent to polar group 3m, 6mm, or 4mm, only the Rashba-like AHNT exists. Besides, the pure Dresselhaus AHNTs can only occur if the symmetry composed on $\boldsymbol{\sigma}_H$ identifies with the non-polar tetragonal point group $\bar{4}$, $\bar{4}$2m, or $\bar{4}$m2. If altermagnetic materials lack higher-order rotational $n$ or helical $n_q$ symmetries (where $n$≥3), there are no constraints on the coefficients of the first-order Néel vector terms. A mixed AHNTs containing both Rashba and Dresselhaus components hence appears (Table I). Under symmetry constraints equivalent to monoclinic (m) or orthogonal (mm2) group, $\boldsymbol{\sigma}_H(\mathbf{n})$ exhibits a mixed Rashba and Dresselhaus AHNTs. Interestingly, when the symmetry associated with $\boldsymbol{\sigma}_H$ corresponds to $\bar{6}$, $\bar{6}$m2, or $\bar{4}$3m point groups, there is no first-order term for $\boldsymbol{\sigma}_H(\mathbf{n})$, resulting in cubic AHNTs as the lowest order in altermagnets. Thus, our classification enables the identification of all AHE behaviors in altermagnets and facilitates the efficient discovery of corresponding materials. Notably, we find a striking similarity between AHNTs and spin textures in momentum space [54,58] (see Supplementary Note 3).

**Table I**. Ten kinds of AHNTs in four categories for altermagnetic materials. The candidate materials are sourced from the MAGNDATA database [59] and Refs. [21,22,35,60].

| AHNT | Expression and equivalent constrained point groups | Candidate materials |
|---|---|---|
| Rashba-like | 3m, 6mm, 4mm<br>$\boldsymbol{\sigma}_H = A(n_y, -n_x, 0)$ | Trigonal: $NiCO_3$, $MnCO_3$, $FeCO_3$, $FeF_3$, $CoF_3$, α-$Fe_2O_3$<br>Tetragonal: $Ta_2FeB_2$, $KMnF_3$ |
| Dresselhaus-like | $\bar{4}$m2: $\boldsymbol{\sigma}_H = A(n_y, n_x, 0)$<br>$\bar{4}$2m: $\boldsymbol{\sigma}_H = B(n_x, -n_y, 0)$<br>$\bar{4}$: $\boldsymbol{\sigma}_H(\bar{4})$<br>$= \boldsymbol{\sigma}_H(\bar{4}\text{m2}) + \boldsymbol{\sigma}_H(\bar{4}\text{2m})$ | Tetragonal: $RuO_2$, $CoF_2$, $MnF_2$, $NiF_2$, $LiFe_2F_6$, $ZrMn_2Ge_4O_{12}$ |
| Mixed texture | m, mm2<br>$\boldsymbol{\sigma}_H = (An_y, Bn_x, 0)$<br>$A \neq \pm B$ | Orthogonal: $LaMnO_3$, $NdMnO_3$, $ScCrO_3$, $NaOsO_3$, $Cu_2V_2O_7$, $NiFePO_5$, $Mn_2SeO_3F_2$ |
| Cubic texture | $\bar{6}$, $\bar{6}$m2 ($\bar{6}$2m), $\bar{4}$3m<br>$\boldsymbol{\sigma}_H = \mathbf{T}^{(4)} : \mathbf{nnn}$ | Hexagonal: MnTe, CrSb, $Ba_3NiRu_2O_9$, $CrNb_4S_8$ |

It should be noted that, similar to the Berry curvature multipoles in moment space [61-63] and the multipoles of spin in real space [39,64-66], Eq. (2) can be viewed as the AHE multipoles in the Néel vector space. In Supplementary Note 5, we project $\boldsymbol{\sigma}_H$ to real sphere harmonics and get the corresponding multipoles in the Néel vector space. The multipoles of $\boldsymbol{\sigma}_H$ in the Néel vector space reflect the complexity of the SOC effect on AHE [8].

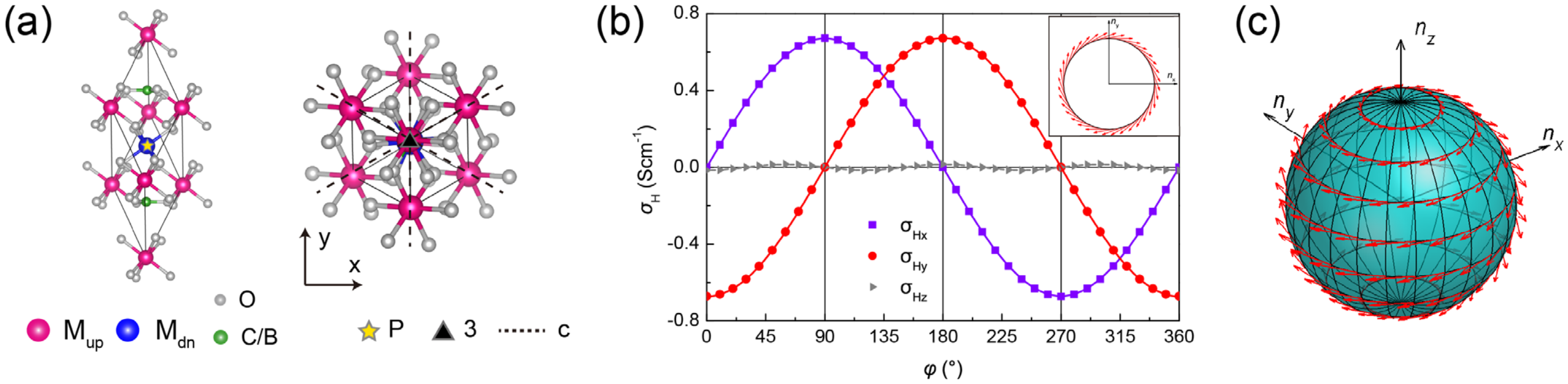

**Fig. 3**. (a) Crystal structure and symmetry of calcite-type altermagnets. (b) Corresponding relations of $\boldsymbol{\sigma}_H(\mathbf{n})$ in the *xoy* plane and (c) AHNT in the whole Néel vector space. The inset in (b) is the Rashba-like AHNT in the xoy plane. The points in (b) are the numerical data calculated by effective TB model (see Supplementary Note 6.3), and the curves are the analytical $\boldsymbol{\sigma}_H(\mathbf{n})$ function fitted by Eq. (7). The parameters of TB model are in Supplementary Note 7.2.

*3.2 Origin of similarity between AHNTs and spin textures.*

To uncover the symmetry origin of the similarity between AHNTs and spin textures, we consider the general SOC Hamiltonian as: $H_{SOC} = \boldsymbol{\Omega}(\mathbf{k}) \cdot \boldsymbol{\sigma}$, where $\boldsymbol{\sigma} = (\sigma_x, \sigma_y, \sigma_z)$ are the Pauli matrices, and $\boldsymbol{\Omega}(\mathbf{k})$ is a wave-vector dependent spin-orbit field [54], which governs the spin expectation at each **k**-point. $\boldsymbol{\Omega}(\mathbf{k})$ behaves as an axial vector, as the AHE conductivity vector $\boldsymbol{\sigma}_H$; whereas, the Néel vector **n** is a polar-like vector as the momentum **k**, because it switches its sign under the inversion symmetry (see Fig. 2 (a)).

The two similarities result in the fact that point group operations impose equivalent constraints on their forms of corresponding expanded coefficients. For a first-order Taylor expansion, $\boldsymbol{\Omega}(\mathbf{k})$can be expressed as:

$$\Omega_i(\mathbf{k}) = \sum_{j=1}^{3} \mathcal{T}_{ij}^{(2)} k_j, \tag{6}$$

where the coefficient matrix $\mathcal{T}_{ij}^{(2)}$ is an axial tensor, similar to $\mathbf{T}^{(2)}$ in Eq. (2). For examples, under the polar 3m, 4mm and 6mm point groups [58], only two nonzero elements of $\mathcal{T}^{(2)}$ exist: $\mathcal{T}_{xy} = -\mathcal{T}_{yx}$ . We can obtain $\boldsymbol{\Omega}(\mathbf{k}) = \mathcal{T}_{xy}(k_y, -k_x, 0)$ and $H_{SOC} = \mathcal{T}_{xy}\left(k_y\sigma_x - k_x\sigma_y\right)$, leading to the Rashba-type spin texture $\mathbf{s}(\mathbf{k})$ ($\mathbf{s} = \frac{\hbar}{2}\langle\varphi|\boldsymbol{\sigma}|\varphi\rangle$, Fig. 1(b)). The configurations of $\mathbf{T}^{(2)}$ tensors align with $\mathcal{T}^{(2)}$ tensors across all point groups, as given in Supplementary Note 3.

Besides, due to the time-reversal symmetry, the spin texture satisfies $\mathbf{s}(\mathbf{k}) = -\mathbf{s}(-\mathbf{k})$ in nonmagnetic materials. A similar relation also exists in AHNTs: $\boldsymbol{\sigma}_{\boldsymbol{H}}(\mathbf{n}) = -\boldsymbol{\sigma}_{\boldsymbol{H}}(-\mathbf{n})$, due to the odd parity of $\boldsymbol{\sigma}_H$ with the time-reversal operation. Consequently, the categories of AHNTs listed in Table I bear a resemblance to the spin texture.

Our theory well reproduces the $\boldsymbol{\sigma}_H(\mathbf{n})$ relation of rutile-type (such as $RuO_2$ [28,30,31,67]) and NiAs-type ($\alpha$-MnTe [25,36]) altermagnetic materials with Dresselhaus-like and pure cubic AHNTs, respectively (see Supplementary Note 6 and 7). The predicted AHNTs are consistent with the AHE [31,36] and X-ray magnetic circular dichroism (XMCD) [68,69] experiments. Except for the two widely studied altermagnets, we introduce two previously unidentified AHNTs in the following.

*3.3 Calcite-type altermagnets*

Altermagnetic materials $NiCO_3$, $MnCO_3$, $FeCO_3$, and $FeBO_3$ have the calcite structure with $R\overline{3}c$ (No. 167) space group. In this structure, the magnetic atoms occupy the 6b positions at (0, 0, 0) and (1/2, 1/2, 1/2), and they are bonded to six equivalent O atoms, forming corner-sharing octahedra, as illustrated in Fig. 3(a). The inversion center is located on each magnetic atom (not the midway point between them), breaking PT symmetry and enabling band splitting even without SOC effect. Unlike the $4_2$ screw symmetry in the rutile-type altermagnets, the three-fold rotational symmetry present here does not switch the opposite-spin magnetic atoms, beside the glide mirror operations interchange them. Consequently, the symmetry composed on $\boldsymbol{\sigma}_H$ identify with 3m1. According to the above symmetry characteristic, we obtain the corresponding $\boldsymbol{\sigma}_H(\mathbf{n})$ (See Supplementary Note 6.3):

$$\boldsymbol{\sigma}_H(\mathbf{n}) = (An_y, -An_x, 0) + O(n^3). \tag{7}$$

Our numerical results for the effective tight-binding (TB) model are consistent with Eq. (7), as shown in Fig. 3 (b). Unlike the rutile-type altermagnets, the $m_z$ mirror symmetry does not exist in calcite altermagnets, which allows $\sigma_{Hz}$ when **n** is in the *xoy* plane and exhibits a third-order contribution $\sigma_{Hz} \sim n_x(n_x^2 - 3n_y^2)$. Although, $\boldsymbol{\sigma}_H(\mathbf{n})$ in *xoy* plane for the rutile-type and calcite-type altermagnets exhibit similar trigonometric characteristics, essential differences leading to calcite-type

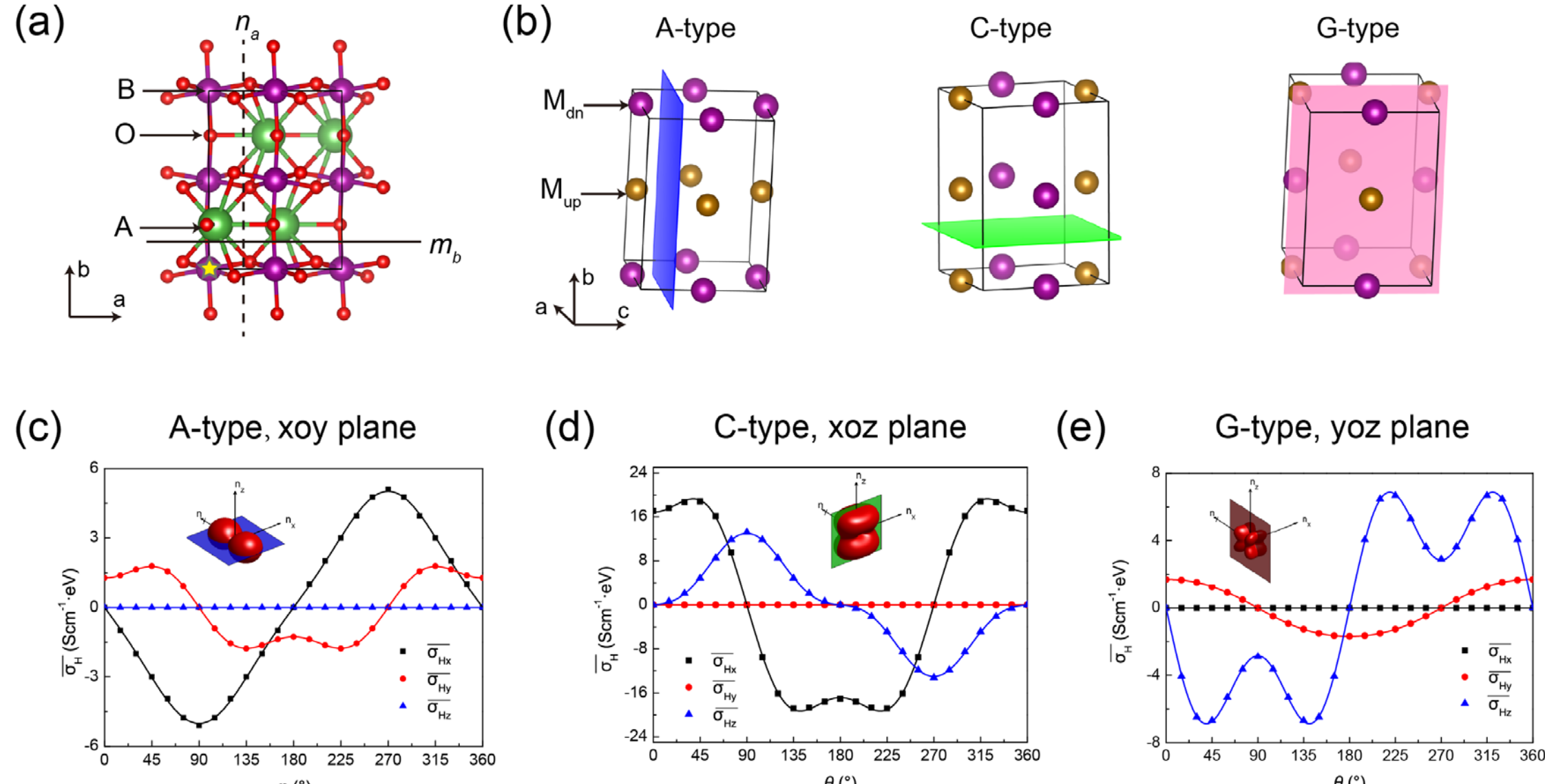

**Fig. 4**. (a) Crystal structure of perovskite-type altermagnetic materials $ABO_3$. (b) Magnetic structure of A-, C- and G-type antiferromagnetic configurations. The magnetic atoms with different spins are represented with different colors. The colored planes indicate that the mirror or glide mirror operations do not switch the magnetic atoms with opposite spins. (c, d, e) Angle dependent of the spectral integrals of optical Hall conductivity $\overline{\boldsymbol{\sigma}_H}$ for A-, C-, and G- type perovskite-type altermagnetic materials, respectively. The solid lines are the fitting curves by Eq. (8). The points in (c-e) are the numerical data calculated by effective TB model (see Supplementary Note 6.4), and the curves are the analytical $\overline{\boldsymbol{\sigma}_H}$ $(\mathbf{n})$ function fitted by Eq. (7). The insects are the abs$|\overline{\boldsymbol{\sigma}_H}|$ profiles in the Néel vector spaces. The parameters of TB model are in altermagnets hold a Rashba-like AHNT when the third order terms are omitted, while the AHNT of rutile-type altermagnets obeys a Dresselhaus-like configuration. This Rashba-like texture in calcite-type altermagnets ensures that the vector $\boldsymbol{\sigma}_H$ is parallel to the latitude plane and tangent to the latitude circles, as shown in Fig. 3(c).

In addition to calcite-type altermagnetic materials, altermagnets with trigonal structures such as $CaLiOsO_6$, $CagLiRuO_6$, $Sr_3LiRuO_6$, $CoF_3$, and $FeF_3$ also exhibit $R\bar{3}c$ symmetry, giving rise to a similar $\boldsymbol{\sigma}_H(\mathbf{n})$ relation. The XMCD [70] and AHE [71] experiments in α-$Fe_2O_3$ show the $\boldsymbol{\sigma}_H \perp \mathbf{n}$ relations when the Néel vector is along the in-plane crystallographic direction directions and $\boldsymbol{\sigma}_H$ vanishes when it is along the [001] direction, suggesting a possible Rashba-like AHNT. In fact, the Rashba-type AHNTs are not limited to trigonal altermagnets. For instance, the tetragonal $KMnF_3$ with the I4/mcm (No. 140) space group and $Ta_2FeB_2$ [35] with the P4/mbm (No. 127) space group can also display Rashba-like AHNTs, as given in Table I.

### *3.4 Perovskite-type altermagnets*

The perovskite-type altermagnetic materials represent a broad class of antiferromagnetic materials in the formula of $ABO_3$, belonging to the space group Pnma (No. 62). An octahedron formed by six oxygen atoms surrounds each magnetic atom B, with non-magnetic A atoms situated at the center of these octahedra, as shown in Fig. 4(a). Typically, the magnetic atoms B occupy the 4b Wyckoff positions at (1/2, 0, 0), (0, 1/2, 1/2), (1/2, 1/2, 0), and (0, 0, 1/2). Consequently, there are 3 ( $C_4^2 C_2^2 / A_2^2 = 3$ ) kinds of collinear antiferromagnetic configurations, indicated by A-type (e.g., $LaMnO_3$, $PrMnO_3$, $NdMnO_3$), C-type (e.g., $CaCrO_3$ [34], $ScCrO_3$), and G-type (e.g., $NaOsO_3$, $LaCrO_3$, $TbCrO_3$, $NdFeO_3$, $TbFeO_3$, etc.), as illustrated in Fig. 4 (b).

Although the symmetry restrictions on $\boldsymbol{\sigma}_H$ are the same in the three magnetic configurations, the constraints on the Néel vector $\mathbf{n}$ turn out to be different (see Supplementary Note 7.4). Therefore, the relations of $\boldsymbol{\sigma}_H(\mathbf{n})$ up to the third-order differ in the three magnetic configurations:

$$\begin{cases} \boldsymbol{\sigma}_H^A(\boldsymbol{n}) = (An_y, Bn_x, 0) + (an_x^2 n_y + bn_y^3 + cn_y n_z^2, dn_x^3 + en_x n_y^2 + fn_x n_z^2, gn_x n_y n_z), \\ \boldsymbol{\sigma}_H^C(\boldsymbol{n}) = (An_z, 0, Bn_x) + (an_x^2 n_z + bn_y^2 n_z + cn_z^3, dn_x n_y n_z, en_x^3 + fn_x n_y^2 + gn_x n_z^2), \\ \boldsymbol{\sigma}_H^G(\boldsymbol{n}) = (0, An_z, Bn_y) + (an_x n_y n_z, bn_x^2 n_z + cn_y^2 n_z + dn_z^3, en_x^2 n_y + fn_y^3 + gn_y n_z^2). \end{cases} \quad (8)$$

The diverse $\boldsymbol{\sigma}_H(\mathbf{n})$ relation is helpful for determining the magnetic structures of perovskite-type altermagnets. It is noteworthy that if the mirror or glide mirror operations do not switch the magnetic atoms with opposite spins (as shown in Fig. 4 (b)), $\boldsymbol{\sigma}_H$ is vanished when the Néel vector $\mathbf{n}$ is perpendicular to the mirror or glide mirror planes, according to Eq. (8). The first order $\boldsymbol{\sigma}_H(\mathbf{n})$ relationship exhibits the mixed Rashba and Dresselhaus AHNTs, because no symmetry forces $A = B$ or $A = -B$.

Since most perovskite-type altermagnetic materials are insulators, we investigate the optical Hall conductivity of A-, C-, and G-type configurations. The calculated results using the TB model are consistent with Eq. (8), as given in Fig. 4 (c-d), where the spectral integrals of optical Hall conductivity $\overline{\boldsymbol{\sigma}_H} = \int_{0+}^{\infty} \boldsymbol{\sigma}_H(\omega)\, \mathrm{d}\omega$ [12,41]. Previous experiments show that some perovskite antiferromagnets [72,73] exhibit the magneto-optical effect, while the $\boldsymbol{\sigma}_H(\mathbf{n})$ relationship in them needs further experiments investigation.

## 4. Conclusions

The extrinsic parameter method is not only applicable to AHE but can also be extended to other magnetically related physical effects [74], such as the second harmonic generation effect [75], the spin Hall effect [76,77], and the nonlinear Hall effect [78-80].

Besides, AHNTs shed light on experimentally detecting the Néel vector in altermagnetic materials via the anomalous Hall transport (See Supplementary Note 9). Importantly, we can utilize the characteristic that the $\boldsymbol{\sigma}_H$ usually is perpendicular to the Néel vector to design a vertical magneto-optical isolator (see Supplementary Note 9).

In summary, we treat the Néel vector as an extrinsic parameter to unveil the relationship between AHE and Néel vector in altermagnets. We present a comprehensive classification of AHNTs for altermagnets, and find a variety of unconventional AHNTs, including Rashba, Dresselhaus, persistent, and pure cubic AHE textures, which are absent in ferromagnetic materials. This complete framework of AHNTs serves as a pictorial category scheme for altermagnets. We found an intriguing similarity between the AHNTs in altermagnets and the spin textures in momentum space, and further elucidated their symmetry origin. Our findings provide a thorough analysis of AHE in altermagnets, and useful guidance for the detection of Néel vectors through AHE, and designing electronic and optoelectronic devices based on altermagnetic materials.

## Competing interest

The authors declare no competing interests.

## Acknowledgments

We appreciate the helpful discussion with S. M. Zhou, Hua Jiang, Yu Dai, Xiao Li, Ze-Ying Zhang, Zhi-Qiang Zhang, Xiao-Dong Zhou, and Jian-Zhou Zhao. This work was supported by the National Key Research and Development Program of China (Grant No. 2024YFA1611300 and 2023YFA1406500), the National Natural Science Foundation of China (Grants Nos. 12474100, 12204009, 12204003, and 12174394), and the Natural Science Foundation of Anhui Province (Grant No. 2208085QA08). J. H. was supported by the HFIPS Director's Fund (Grants No. YZJJQY202304 and No. BJPY2023B05) and Anhui Provincial Major S&T Project (s202305a12020005), the High Magnetic Field Laboratory of Anhui Province under Contract No. AHHM-FX-2020-02 and the Basic Research Program of the Chinese Academy of Sciences Based on Major Scientific Infrastructures No. JZHKYPT-2021-08. R.-C.X. acknowledges the startup foundation from AHU.

## Author contributions

R.C.X. conceived the project and edited the code and performed the calculations. R.C.X. performed the symmetry and TB model analysis, and J.H.Z. checked it. R.C.X., J.H.Z., and M.L.T. discussed the results and the writing. The manuscript was written through the contributions of all authors. All authors have approved the final version of the manuscript.

## Appendix A. Supplementary materials

Supplementary materials to this article can be found online at https:/xxxxx.

SUPPLEMENTARY MATERIALS

# Anomalous-Hall Néel textures in altermagnetic materials

Rui-Chun Xiao,[1, 2, *] Hui Li,[1, 2] Hui Han,[1,2] Wei Gan,[1] Mengmeng Yang,[1, 2] Ding-Fu Shao,[3] Shu-Hui Zhang,[4] Yang Gao,[5] Mingliang Tian,[6, 7, †] Jianhui Zhou[6, ‡]

[1] *Institute of Physical Science and Information Technology, Anhui University, Hefei 230601, China*

[2]*Anhui Provincial Key Laboratory of Magnetic Functional Materials and Devices, School of Materials Science and Engineering, Anhui University, Hefei 230601, China*

[3]*Key Laboratory of Materials Physics, Institute of Solid State Physics, Chinese Academy of Sciences, 230031 Hefei, China*

[4]*College of Mathematics and Physics, Beijing University of Chemical Technology, Beijing 100029, China*

[5]*Department of Physics, University of Science and Technology of China, Hefei 230026, China.*

[6]*Anhui Key Laboratory of Low-Energy Quantum Materials and Devices, High Magnetic Field Laboratory, HFIPS, Anhui, Chinese Academy of Sciences, Hefei, Anhui 230031, China.*

[7]*School of Physics and Optoelectronics Engineering, Anhui University, Hefei 230601, China*

[*]xiaoruichun@ahu.edu.cn; [†]mltian@ahu.edu.cn; [‡]jhzhou@hmfl.ac.cn

## Supplementary Note 1. Transformations of Néel vector under the space group operations

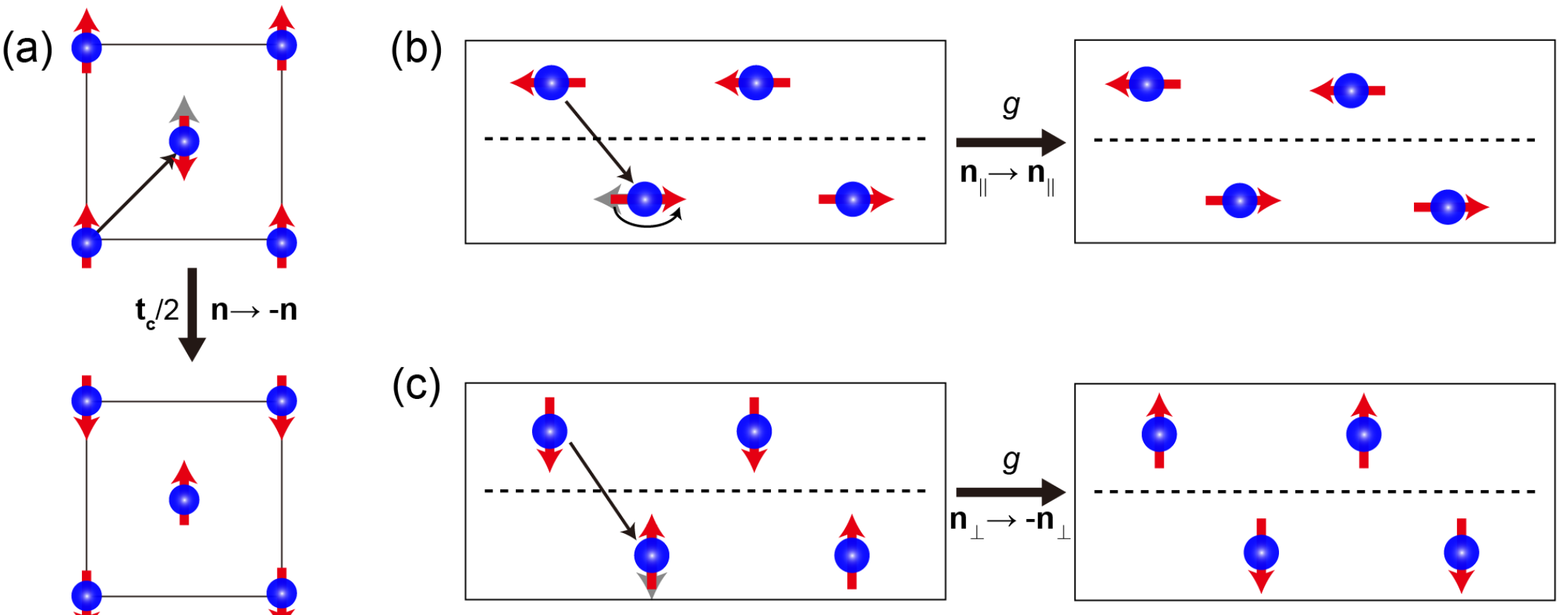

Fig. S1. (a) Transformation of Néel vector under the $\mathbf{t}_c/2$ translation operation. (b) Transformations of the (b) horizontal and (c) vertical components of Néel vectors under the glide mirror operation.

The transformations of the Néel vector under the space group operation are illustrated in Fig. S1. The spin remains invariant under the translation operation, therefore $\mathbf{t}_c/2$ ($\mathbf{t}_c$ equals the integral lattice vectors) operation, which connects magnetic atoms with opposite spins, reverses the Néel vectors, as shown in Fig. S1(a). As an axial vector, the spin reverses when parallel to the glide mirror (Fig. S1(b)), and remains invariant when perpendicular to the glide mirror (Fig. S1(c)), resulting in the invariant and reversal of Néel vectors, respectively.

## Supplementary Note 2. Extrinsic parameter method and algorithm

First, we investigate the relationship between the anomalous Hall conductivity vector $\boldsymbol{\sigma}_H$ and the Néel vector $\mathbf{n}$ at first-order Taylor approximation. Under this situation, $\boldsymbol{\sigma}_H$ can be expanded in the following form:

$$\begin{cases} \sigma_{Hx} = T_{xx}n_x + T_{xy}n_x + T_{xz}n_z, \\ \sigma_{Hy} = T_{yx}n_x + T_{yy}n_x + T_{yz}n_z, \\ \sigma_{Hz} = T_{zx}n_x + T_{zy}n_x + T_{zz}n_z. \end{cases} \tag{1}$$

This indicates that the anomalous Hall vector $\boldsymbol{\sigma}_H$ and the Néel vector $\mathbf{n}$ are connected by a 3×3 tensor, expressed as:

$$\sigma_{Hi} = \left(\mathbf{T}^{(2)} \cdot \mathbf{n}\right)_i = \sum_j T_{ij}^{(2)} n_j. \tag{2}$$

Here, $\mathbf{T}^{(2)}$ is a two-dimensional tensor whose elements are in Eq. (1), *i.e.*,

$$\mathbf{T}^{(2)} = \begin{bmatrix} T_{xx} & T_{xy} & T_{xz} \\ T_{yx} & T_{yy} & T_{yz} \\ T_{zx} & T_{zy} & T_{zz} \end{bmatrix}. \tag{3}$$

Under the space group operation $\{R\,|\,\mathbf{t}\}$, $\boldsymbol{\sigma}_H$ and $\mathbf{n}$ are transformed into $\boldsymbol{\sigma}'_H$ and $\mathbf{n}'$, which obey the following relationship:

$$\boldsymbol{\sigma}'_H = \det(R) D(R) \boldsymbol{\sigma}_H, \tag{4}$$

$$\mathbf{n}' = \pm \det(R) D(R) \mathbf{n}, \tag{5}$$

where $D(R)$ is the matrix form of point symmetry operation $R$. If $\{R\,|\,\mathbf{t}\}$ does not exchange magnetic atoms with opposite spins, we take the positive sign in Eq. (5); if $\{R|\mathbf{t}\}$ switches magnetic atoms with opposite spins, we take the negative sign. This can be expressed as:

$$\{R\,|\,\mathbf{t}\}\mathbf{r}_{i,\uparrow} = R\mathbf{r}_{i,\uparrow} + \mathbf{t} \begin{cases} = \mathbf{r}_{j,\uparrow} + (n_1\mathbf{a} + n_2\mathbf{b} + n_3\mathbf{c}), & +1 \\ = \mathbf{r}_{j,\downarrow} + (n_1\mathbf{a} + n_2\mathbf{b} + n_3\mathbf{c}), & -1 \end{cases} \tag{6}$$

where $\mathbf{a}$, $\mathbf{b}$, and $\mathbf{c}$ are the lattice vectors in three directions, $n_1$, $n_2$, and $n_3$ are integers.

Similar to Eq. (2), the transformed vectors $\boldsymbol{\sigma}'_H$ and $\mathbf{n}'$ satisfy the relationship: $\boldsymbol{\sigma}'_H = \mathbf{T}'^{(2)} \cdot \mathbf{n}'$. Thus, $\mathbf{T}'^{(2)}$ and $\mathbf{T}^{(2)}$ must satisfy the following transformation relationship:

$$\mathbf{T}'^{(2)} = \pm D(R)\mathbf{T}^{(2)} D(R^{-1}). \tag{7}$$

In Eq. (7), $\det(R)$ is eliminated because both sides of the equation all have axial vectors. Furthermore, Eq. (7) can also be expressed in summation form

$$T_{ij}'^{(2)} = \pm \sum_{mn} R_{im} R_{jn} T_{mn}^{(2)}, \tag{8}$$

where $R_{ij}$ is an element of the matrix $D(R)$.

If the anomalous Hall vector $\boldsymbol{\sigma}_H$ is further expanded to the third-order terms of the Néel vector $\mathbf{n}$, similar to Eq. (2), it can be expressed with a fourth-order tensor:

$$\boldsymbol{\sigma}_H = \mathbf{T}^{(4)} \vdots \mathbf{nnn}. \tag{9}$$

Analogous to Eq. (8), the transformation of the fourth-dimensional tensor $\mathbf{T}^{(4)}$ under symmetry operations is given by:

$$T^{(4)}_{ijkl} = \pm R_{im} R_{jn} R_{kp} R_{lq} T^{(4)}_{mnpq}. \tag{10}$$

Additionally, since the expression in Eq. (9) involves the multiplication of three Néel vectors, $\mathbf{T}^{(4)}$ exhibits commutative symmetry for the last three indices:

$$T^{(4)}_{ijkl} = T^{(4)}_{ijlk} = T^{(4)}_{ikjl} = T^{(4)}_{iklj} = T^{(4)}_{ilkj} = T^{(4)}_{iljk}. \tag{11}$$

As a result, the number of independent tensor elements of $\mathbf{T}^{(4)}$ decreases from $3^4$=81 to 30.

Considering both the first-order and third-order Taylor expansions of $\mathbf{n}$, the $\boldsymbol{\sigma}_H$ can ultimately be expressed as:

$$\sigma_{Hi}(\mathbf{n}) = \sum_{j=1}^{3} T^{(2)}_{ij} n_j + \sum_{j,k,l=1}^{3} T^{(4)}_{ijkl} n_j n_k n_l. \tag{12}$$

```
Clear["Global`*"]
Get[NotebookDirectory[] <> "\\Script\\" <> "SGData.wl"]
Get[NotebookDirectory[] <> "\\Script\\" <> "AHENeel.wl"]

{SGop, Basis} = SPaceGroupData[194]; (*NiAs type*)
Magup = {{0, 0, 0}};   (*Magnetic atom postion with spin up state,in crystal coordinate*)
Magdn = {{0, 0, 1/2}}; (*Magnetic atom postion with spin down state,in crystal coordinate*)
sigma = AHENeel[SGop, Basis, Magup];

Space group name: P6_3/mmc
Bravais lattice type: HexagonalP
Conventional Lattice Vectors: {{a, 0, 0}, {-a/2, √3 a/2, 0}, {0, 0, c}}
Primitive Lattice Vectors: {{a, 0, 0}, {-a/2, √3 a/2, 0}, {0, 0, c}}
```

$$\sigma_H = \begin{pmatrix} 0 \\ 0 \\ 0 \end{pmatrix} + \begin{pmatrix} 6\, n_x n_y n_z T_{Xxyz} \\ 3 n_x^2 n_z T_{Xxyz} - 3 n_y^2 n_z T_{Xxyz} \\ 3 n_x^2 n_y T_{Zxxy} - n_y^3 T_{Zxxy} \end{pmatrix}$$

```
Independent 1st order terms: {}, Number:0
Independent 3rd order terms: {Txxyz, Tzxxy}, Number:2
```

Fig. S2. Usage of Mathematica code “AHE-texture” to determine the $\boldsymbol{\sigma}_H(\mathbf{n})$ relation.

Based on Eq. (8) and Eq. (10), we have developed the corresponding computational code “AHE-texture” based on Wolfram Mathematica, which was shared on GitHub (https://github.com/Ruichun/AHE-texture). The usage of code is shown in Fig. S2. Firstly, we load the code, then input the space group number and the positions of the magnetic atoms with spin-up and spin-down states. After running the codes, the Taylor expansion polynomials of anomalous Hall vector $\boldsymbol{\sigma}_H$ with $\mathbf{n}$, as well as the independent tensor elements, will be outputted automatically.

Although our research focuses on altermagnets, the proposed method is also suitable for ferromagnets. To illustrate this point, we have successfully derived the anomalous Hall conductivity for Fe (space group $Im\overline{3}m$ ) as a function of its magnetic moment:

$$\boldsymbol{\sigma}_H = \begin{pmatrix} T_{Xx}m_x \\ T_{Xx}m_y \\ T_{Xx}m_z \end{pmatrix} + \begin{pmatrix} T_{Xxxx}m_x^3 + 3T_{Xxyy}m_x m_y^2 + 3T_{Xxyy}m_x m_z^2 \\ T_{Xxxx}m_y^3 + 3T_{Xxyy}m_x^2 m_y + 3T_{Xxyy}m_z^2 m_y \\ T_{Xxxx}m_z^3 + 3T_{Xxyy}m_x^2 m_z + 3T_{Xxyy}m_y^2 m_z \end{pmatrix} = A\begin{pmatrix} m_x \\ m_y \\ m_z \end{pmatrix} + B\begin{pmatrix} m_x^3 \\ m_y^3 \\ m_z^3 \end{pmatrix}, \quad (13)$$

where $A = T_{Xx} + 3T_{Xxyy}$ , $B = T_{Xxxx} - 3T_{Xxyy}$ , $\mathbf{m} = \left(m_x, m_y, m_z\right)$ is the magmatic moment vector with $m_x^2 + m_y^2 + m_z^2 = 1$ . The above result is well consistent with previously reported outcomes in Refs. [1-3].

The magnetic group approach considers magnetic order parameters as intrinsic structural parameters, as given in Fig. S3(a). The magnetic group methods are limited to determine the direction of $\boldsymbol{\sigma}_H$ for specific crystalline orientations (see Supplementary Note 8). This method mainly determines the $\boldsymbol{\sigma}_H$ direction in specific crystallographic directions (e.g., [100], [010], [001], etc.), making it challenging to establish an analytical $\boldsymbol{\sigma}_H - \mathbf{n}$ relationship. Readers can refer to the MTENSOR module in Bilbao Crystallographic Server (https://www.cryst.ehu.es/cgi-bin/cryst/programs/mtensor.pl, and the corresponding theory presented Ref. [4] ) for more information.

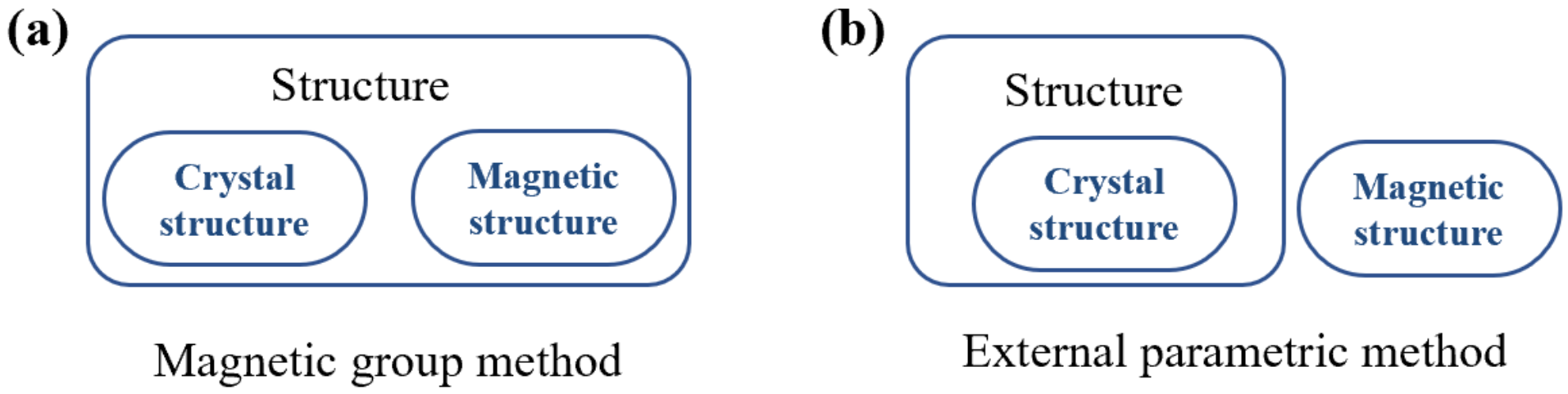

Fig. S3 Illustration the differences between (a) the magnetic group method and (b) the external parameter method in treating the magnetic order.

Unlike the traditional magnetic group method that treats both the magnetic moment and crystal structure as intrinsic parameters, we instead analyze the Néel vector as an extrinsic parameter, as illustrated in Fig. S3 (b). In contrast, the extrinsic parameter method offers a broader capability, enabling the establishment of relationships among different magnetic orientations. Besides, the "extrinsic parameter" method only considers the symmetry operations of space groups. Although this method does not directly involve time-reversal symmetry operations like the magnetic groups, it is noteworthy that the ±1 symbol in Eq. (5) essentially reflects the sign change of the magnetic moment under time-reversal operation.

Besides, the linear-response-symmetry code (https://bitbucket.org/zeleznyj/linear-response-symmetry/src/master/, and corresponding theory presented in Ref. [5]) is the pioneering work on determining various symmetry properties of magnetic and nonmagnetic crystals. With the aid of the site point group symmetry, this approach enables us to derive the response tensors through a Taylor expansion of the magnetic moment. This method demonstrates remarkable efficiency in determining the response tensors for ferromagnetic systems. On the other hand, "*in a two-sublattice collinear antiferromagnet, if a projection on one of the magnetic atoms is specified, then this expansion also applies. In general, however, for a collinear antiferromagnet the expansion is different from a ferromagnet and this is not implemented!*", as stated on its official website. This unimplemented functionality for antiferromagnetic materials in the linear-response-symmetry code calls for effective and convenient methods to handle the transformation of the magnetic order parameters under symmetry operations.

The comparison of linear-response-symmetry, extrinsic parameter method and magnetic group method is summarized in Table SI.

Table SI. The comparison of linear-response-symmetry, extrinsic parameter method, and Bilbao Crystallographic Server.

| | **Linear-response-symmetry method** | **Extrinsic parameter method** | **Magnetic group method** |
|---|---|---|---|
| Group | Site point group | Space group | Magnetic group |
| Suitable materials | Ferromagnets or one sublattice in antiferromagnets | Ferromagnets and (collinear) antiferromagnets | All the magnets |
| Order parameter | Magnetic moment | Magnetic moment or Néel vector | × Not available |
| $\boldsymbol{\sigma}_H - \mathbf{n}$ analytical relation | Not available now | √ | ×, can only determine the direction of $\boldsymbol{\sigma}_H$ for special n direction. |

## Supplementary Note 3. The comparison of spin texture and anomalous Hall texture

In nonmagnetic materials, the combination of the SOC effect and the breaking of inversion symmetry induces spin splitting of electronic bands, leading to the spin texture in momentum space. Similarly, AHE arises from SOC effect [3] and time-reversal symmetry breaking.

The SOC Hamiltonian is given by $H_{SOC} = \Omega(\mathbf{k}) \cdot \boldsymbol{\sigma}$, where $\boldsymbol{\sigma} = (\sigma_x, \sigma_y, \sigma_z)$ are the Pauli matrices, and $\boldsymbol{\Omega}(\mathbf{k})$ is a wave-vector dependent spin-orbit field [6]. $\boldsymbol{\Omega}(\mathbf{k})$ governs the spin at each $\mathbf{k}$-point. $\Omega(\mathbf{k})$ behaves as a pseudovector, as the AHE conductivity vector $\boldsymbol{\sigma}_H$. For a first-order Taylor expansion of wave vector $\mathbf{k}$, $\Omega(\mathbf{k})$ can be expressed as:

$$\Omega_i(\mathbf{k}) = \sum_{j=1}^{3} \mathcal{T}_{ij}^{(2)} k_j. \tag{14}$$

where the 1st Taylor expansion coefficient matrix $\left[\mathcal{T}_{ij}^{(2)}\right]$ forms an axial tensor. The symmetry constraining on any axial tensor is the same:

$$\mathcal{T}^{(2)} = \det(R) D(R) \mathcal{T}^{(2)} D(R^{-1}). \tag{15}$$

For example, under the polar 3m, 4mm, and 6mm point groups, we can obtain the symmetry-constrained $\mathcal{T}^{(2)}$ based on Eq. (15) as follows

$$\left.\begin{matrix} 3m \\ 4mm \\ 6mm \end{matrix}\right\} : \mathcal{T}^{(2)} = \begin{pmatrix} 0 & \mathcal{T}_{xy} & 0 \\ -\mathcal{T}_{xy} & 0 & 0 \\ 0 & 0 & 0 \end{pmatrix}. \tag{16}$$

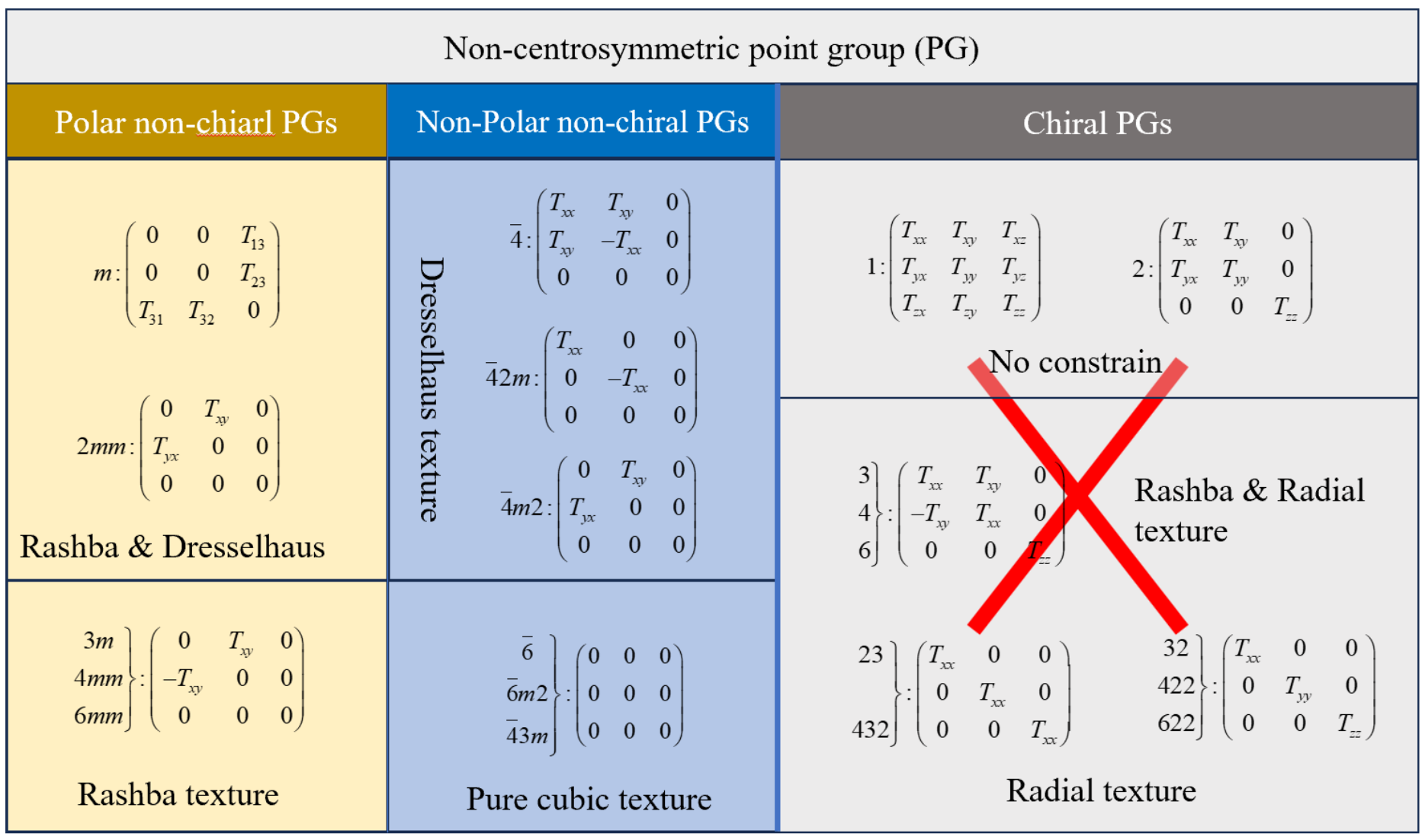

Fig. S4. The equivalent point group symmetry constrained of $\mathbf{T}^{(2)}$ and corresponding AHE texture in 21 noncentrosymmetric point groups. The forms of $\mathbf{T}^{(2)}$ and associated AHTNs are the same as the $\mathcal{T}^{(2)}$ and its spin texture. The equivalent chiral point groups are not allowed for AHNTs in altermagnets due to the uncompensated magnetic moments.

Putting Eq. (16) to (14), we can obtain the SOC vector field $\Omega(\mathbf{k}) = \lambda_R(-k_y, k_x, 0)$ ($\lambda_R = -\mathcal{T}_{xy}$), consequently $H_{SOC} = \lambda_R\left(k_x\sigma_y - k_y\sigma_x\right)$, leading to the Rashba-type spin texture. Similarly, under the chiral point group $\bar{4}m2$, the expression of $\mathcal{T}^{(2)}$ is

$$\bar{4}m2 : \mathcal{T}^{(2)} = \begin{pmatrix} 0 & \mathcal{T}_{xy} & 0 \\ \mathcal{T}_{xy} & 0 & 0 \\ 0 & 0 & 0 \end{pmatrix}. \tag{17}$$

Putting Eq. (17) to (14), we can obtain $\Omega(\boldsymbol{k}) = \beta_D(k_y, k_x, 0)$ ($\beta_D = \mathcal{T}_{xy}$), consequently $H_{SOC} = \beta_D(k_x\sigma_y + k_y\sigma_x)$, leading to the Desslhaus-type spin texture. Under the $\overline{6}$, $\overline{6}$m2 ($\overline{6}$2m), and $\overline{4}$3m point groups, the existence of $\mathcal{T}^{(2)}$ is not allowed. Therefore, there is no first-order term of $\Omega(\mathbf{k})$, resulting in pure cubic spin texture. The above results are consistent with Ref. [7]. The results on $\mathcal{T}^{(2)}$ presented above resemble the outcomes depicted in the AHE texture on $\mathbf{T}^{(2)}$, as given in Fig. S4

The comparison of spin texture and AHNTs is presented in Table S2. The spin $\mathbf{s}$ and anomalous Hall conductivity $\boldsymbol{\sigma}_H$ axial vectors. Similar to moment $\mathbf{k}$, the Néel vector is a polar-like vector, because it switches its sign under the inversion symmetry (see Fig. 2 (a) of the main text). Due to the time-reversal symmetry, the spin texture satisfies $\mathbf{s}(\mathbf{k}) = -\mathbf{s}(-\mathbf{k})$ in nonmagnetic materials. A similar relation also exists in anomalous Hall texture: $\boldsymbol{\sigma}_{\boldsymbol{H}}(\mathbf{n}) = -\boldsymbol{\sigma}_{\boldsymbol{H}}(-\mathbf{n})$, due to the odd parity of $\boldsymbol{\sigma}_H$ with the time-reversal operation. In all, the anomalous Hall vector $\boldsymbol{\sigma}_H$ in the Néel vector space closely resembles the spin in the moment space. This similarity can be attributed to the fact that both the spin-orbit coupling (SOC) field and the AHE vector $\boldsymbol{\sigma}_H$ are pseudovectors, and they are governed by similar (equivalent) point group symmetries. By drawing an analogy to the spin texture, we can refer to the $\boldsymbol{\sigma}_H - \mathbf{n}$ relation as the anomalous Hall texture in the Néel vector space.

Table S2. Comparison of spin texture in nonmagnetic materials and anomalous Hall texture in altermagnets.

| | **Spin texture** | **anomalous Hall texture** |
|---|---|---|
| Relationship | Spin-wave vector<br>$\mathbf{s} - \mathbf{k}$ | Anomalous Hall vector-Néel vector<br>$\boldsymbol{\sigma}_{\mathrm{H}} - \mathbf{n}$ |
| Vector types | $\mathbf{s}$: axial vector<br>$\mathbf{k}$: polar vector | $\boldsymbol{\sigma}_H$: axial vector<br>$\mathbf{n}$: polar-like vector |
| Conditions of emergence | SOC effect, spatial inversion symmetry breaking | SOC effect, time-reversal symmetry breaking |
| Time-reversal operation | $\mathbf{s}(\mathbf{k}) = -\mathbf{s}(-\mathbf{k})$ | $\boldsymbol{\sigma}_{\boldsymbol{H}}(\boldsymbol{n}) = -\boldsymbol{\sigma}_{\boldsymbol{H}}(-\mathbf{n})$ |

## Supplementary Note 4. Completeness of $\boldsymbol{\sigma}_{\boldsymbol{H}}(\mathbf{n})$ relations in altermagnets

As mentioned above, we can transform the $\boldsymbol{\sigma}_H - \mathbf{n}$ relations within the context of magnetic groups into a problem related to nonmagnetic space groups. It is important to note that in Eqs. (8) and (10), aside from the ±1 sign, the remaining components are equivalent to the point group operations $R$ acting on second- and fourth-order tensors. If we interpret the ±1 arising from the exchange of magnetic atoms in these equations as the determinant of the point operation matrix $R$ (det($R$)), we can see that $\mathbf{T}^{(2)}$ and $\mathbf{T}^{(4)}$ exhibit similarities to the second- and fourth-order axial tensor, respectively. Specifically, when $\{R|\mathbf{t}\}$ operations exchange magnetic atoms, we define the determinant of $R$ as negative, and treat $R$ as improper point operations $IR$ ($I$ is the inversion operation) for the axial tensors $\mathbf{T}^{(2)}$ and $\mathbf{T}^{(4)}$. Conversely, if the magnetic atoms are not exchanged under $\{R|\mathbf{t}\}$, we set det($R$)=1, and $R$ acts as the proper point operations on $\mathbf{T}^{(2)}$ and $\mathbf{T}^{(4)}$. Therefore, we can use the equivalent point group (PG) to seek all the possible forms of the AHNTs.

Thus, solving all $\boldsymbol{\sigma}_{\boldsymbol{H}}(\mathbf{n})$ relationships is equivalent to exploring the behavior of second- and fourth-order axial tensors under the 32 crystallographic point groups. We can leverage group theory to simplify this exploration and investigate the completed $\boldsymbol{\sigma}_{\boldsymbol{H}}(\mathbf{n})$ relations in altermagnets. Specifically, the AHE tensors $\mathbf{T}^{(2)}$ exhibit similarities with general second-order axial tensors. In the context of the "extrinsic parameter" method, it suffices to utilize the nonmagnetic group, rather than the magnetic group, to investigate the symmetry constraints imposed on $\mathbf{T}^{(2)}$.

Because the inversion symmetry prohibits the presence of axial tensor elements, only 21 non-centrosymmetric crystallographic point groups can accommodate nonzero second- and fourth-order axial tensors. We list all the equivalent symmetry-constrained axial tensors $\mathbf{T}^{(2)}$ as shown in Fig. S4, where the corresponding AHE textures are also listed.

For fully compensated collinear antiferromagnets, there must exist a symmetry to connect two magnetic atoms with opposite spins. For this case, it is equivalent to constraining that there must exist an R with det(R)=-1 for an equivalent point group in Eq. (8) and (10). While for the chiral groups, there is no symmetry that makes the magnetic moment compensated, and they are also ruled out. 11 chiral point groups (1, 2, 222, 4, 422, 3, 32, 6, 622, 23, and 432) lack improper symmetry operations with a determinant of det($R$)=−1.

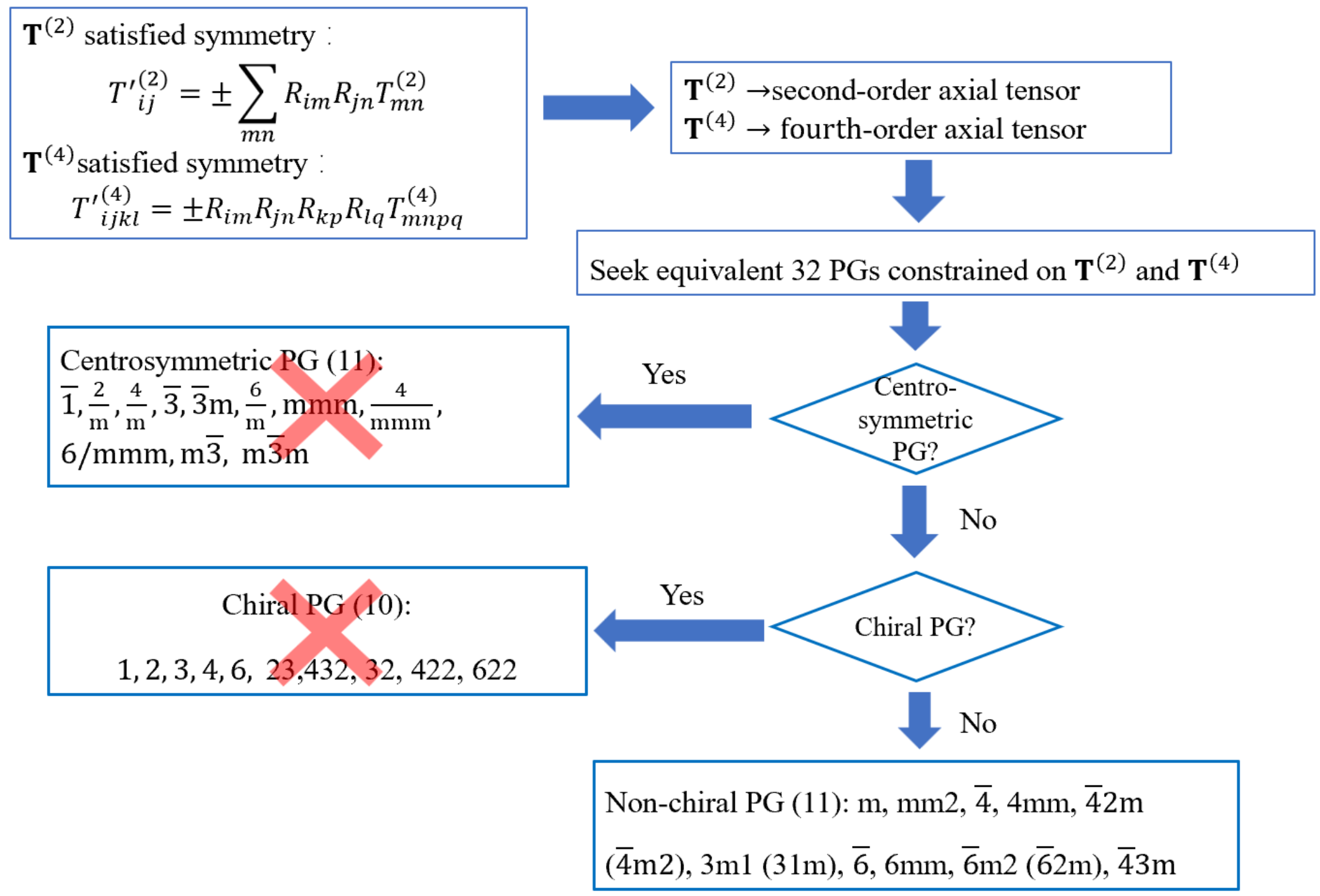

Fig. S5 Flow chart to determine all the AHNTs using the equivalent point groups (PG).

Based on these constraints, we identified 10 non-centrosymmetric and non-chiral point groups (m, mm2, $\bar{4}$, 4mm, $\bar{4}$2m ($\bar{4}$m2), 3m1 (31m), $\bar{6}$, 6mm, $\bar{6}$m2 ($\bar{6}$2m), and $\bar{4}$3m) to explore all possible cases of the $\boldsymbol{\sigma}_H(\mathbf{n})$ relations in altermagnets. The flow chart of the above progress is shown in Fig. S5. As summarized in Table S3, the $\boldsymbol{\sigma}_H(\mathbf{n})$ relations exhibit the following characteristics:

1. There are no equal diagonal terms for $\mathbf{T}^{(2)}$ in all the above point groups, which indicates that altermagnetic materials do not allow for radial AHNTs, such as $A(n_x, n_y, 0)$ or $A(n_x, n_y, n_z)$.
2. In the $\bar{6}$, $\bar{6}$m2 ($\bar{6}$2m), and $\bar{4}$3m point groups, the existence of second-order axial tensors $\mathbf{T}^{(2)}$ is not allowed. Therefore, there is no first-order term for $\boldsymbol{\sigma}_H(\mathbf{n})$, resulting in pure cubic AHNTs in altermagnets.
3. If the symmetry composed on $\boldsymbol{\sigma}_H$ is equivalent to the 3m, 4mm, or 6mm, only the Rashba term $\boldsymbol{\sigma}_H = A(n_y, -n_x, 0)$ exists for the first-order term in AHNTs.
4. The pure Dresselhaus AHNTs can only occur if the symmetry composed on $\boldsymbol{\sigma}_H$ identifies with the tetragonal group $\bar{4}$2m ($\boldsymbol{\sigma}_H = A(n_y, n_x, 0)$), $\bar{4}$m2 ($\boldsymbol{\sigma}_H = A(n_x, -n_y, 0)$), or $\bar{4}$ ($\boldsymbol{\sigma}_H = A(n_y, n_x, 0) + B(n_x, -n_y, 0)$).

5. Under the symmetry constraints equivalent to the monoclinic (m) or orthogonal (mm2) group, $\boldsymbol{\sigma}_H(\mathbf{n})$ exhibits a mixed Rashba and Dresselhaus texture.

Table S3. All the possible $\boldsymbol{\sigma}_H(\mathbf{n})$ relations, and corresponding AHNTs based on 10 non-centrosymmetric and non-chiral point groups.

| **Point group** | **Expressions** | **Independent elements** | **AHNTs** |
| --- | --- | --- | --- |
| m<br>(m⊥z) | $\boldsymbol{\sigma}_H(\mathbf{n})=\begin{pmatrix} T_{\mathrm{Xz}}n_z \\ T_{\mathrm{Yz}}n_z \\ T_{\mathrm{Zx}}n_x+T_{\mathrm{Zy}}n_y \end{pmatrix}+\begin{pmatrix} T_{\mathrm{Xzzz}}n_z^3+3n_x^2T_{\mathrm{Xxxz}}n_z+6T_{\mathrm{Xxyz}}n_xn_yn_z+3T_{\mathrm{Xyyz}}n_y^2n_z \\ T_{\mathrm{Yzzz}}n_z^3+3n_x^2T_{\mathrm{Yxxz}}n_z+6n_xn_yT_{\mathrm{Yxyz}}n_z+3T_{\mathrm{Yyyz}}n_y^2n_z \\ T_{\mathrm{Zxxx}}n_x^3+3T_{\mathrm{Zxxy}}n_yn_x^2+3T_{\mathrm{Zxyy}}n_y^2n_x+3n_z^2T_{\mathrm{Zxzz}}n_x+T_{\mathrm{Zyyy}}n_y^3+3T_{\mathrm{Zyzz}}n_yn_z^2 \end{pmatrix}$ | 1st order: 4<br>3rd order: 14 | Mixed Rashba and Dresselhaus texture |
| mm2<br>(2//z) | $\boldsymbol{\sigma}_H(\mathbf{n})=\begin{pmatrix} T_{\mathrm{Xy}}n_y \\ T_{\mathrm{Yx}}n_x \\ 0 \end{pmatrix}+\begin{pmatrix} T_{\mathrm{Xyyy}}n_y^3+3T_{\mathrm{Xxxy}}n_x^2n_y+3T_{\mathrm{Xyzz}}n_z^2n_y \\ T_{\mathrm{Yxxx}}n_x^3+3T_{\mathrm{Yxyy}}n_y^2n_x+3T_{\mathrm{Yxzz}}n_z^2n_x \\ 6T_{\mathrm{Zxyz}}n_xn_yn_z \end{pmatrix}$ | 1st order: 2<br>3rd order: 7 | Mixed Rashba and Dresselhaus texture |
| $\bar{4}$ | $\boldsymbol{\sigma}_H(\mathbf{n})=\begin{pmatrix} T_{\mathrm{Xx}}n_x+T_{\mathrm{Xy}}n_y \\ T_{\mathrm{Xy}}n_x-T_{\mathrm{Xx}}n_y \\ 0 \end{pmatrix}+\begin{pmatrix} T_{\mathrm{Xxxx}}n_x^3+3T_{\mathrm{Xxxy}}n_yn_x^2+3T_{\mathrm{Xxyy}}n_y^2n_x+3T_{\mathrm{Xxzz}}n_z^2n_x+T_{\mathrm{Xyyy}}n_y^3+3T_{\mathrm{Xyzz}}n_yn_z^2 \\ T_{\mathrm{Xyyy}}n_x^3-3T_{\mathrm{Xxyy}}n_yn_x^2+3T_{\mathrm{Xxxy}}n_y^2n_x+3T_{\mathrm{Xyzz}}n_z^2n_x-T_{\mathrm{Xxxx}}n_y^3-3T_{\mathrm{Xxzz}}n_yn_z^2 \\ 3T_{\mathrm{Zxxz}}n_zn_x^2+6T_{\mathrm{Zxyz}}n_yn_zn_x-3T_{\mathrm{Zxxz}}n_y^2n_z \end{pmatrix}$ | 1st order: 2<br>3rd order: 8 | Dresselhaus |
| 4mm | $\boldsymbol{\sigma}_H(\mathbf{n})=\begin{pmatrix} T_{\mathrm{Xy}}n_y \\ -T_{\mathrm{Xy}}n_x \\ 0 \end{pmatrix}+\begin{pmatrix} T_{\mathrm{Xyyy}}n_y^3+3T_{\mathrm{Xxxy}}n_x^2n_y+3T_{\mathrm{Xyzz}}n_z^2n_y \\ -T_{\mathrm{Xyyy}}n_x^3-3T_{\mathrm{Xxxy}}n_y^2n_x-3T_{\mathrm{Xyzz}}n_z^2n_x \\ 0 \end{pmatrix}$ | 1st order: 1<br>3rd order: 3 | Rashba |
| $\bar{4}2m$ | $\boldsymbol{\sigma}_H(\mathbf{n})=\begin{pmatrix} T_{\mathrm{Xx}}n_x \\ -T_{\mathrm{Xx}}n_y \\ 0 \end{pmatrix}+\begin{pmatrix} T_{\mathrm{Xxxx}}n_x^3+3T_{\mathrm{Xxyy}}n_xn_y^2+3T_{\mathrm{Xxzz}}n_xn_z^2 \\ -3T_{\mathrm{Xxyy}}n_x^2n_y-T_{\mathrm{Xxxx}}n_y^3-3T_{\mathrm{Xxzz}}n_yn_z^2 \\ 3T_{\mathrm{Zxxz}}n_x^2n_z-3T_{\mathrm{Zxxz}}n_y^2n_z \end{pmatrix}$ | 1st order: 1<br>3rd order: 4 | Dresselhaus |
| $\bar{4}m2$ | $\boldsymbol{\sigma}_H(\mathbf{n})=\begin{pmatrix} n_yT_{\mathrm{Xy}} \\ n_xT_{\mathrm{Xy}} \\ 0 \end{pmatrix}+\begin{pmatrix} T_{\mathrm{Xyyy}}n_y^3+3T_{\mathrm{Xxxy}}n_x^2n_y+3T_{\mathrm{Xyzz}}n_z^2n_y \\ T_{\mathrm{Xyyy}}n_x^3+3T_{\mathrm{Xxxy}}n_y^2n_x+3T_{\mathrm{Xyzz}}n_z^2n_x \\ 6T_{\mathrm{Zxyz}}n_xn_yn_z \end{pmatrix}$ | 1st order: 1<br>3rd order: 4 | Dresselhaus |
| 3m1 | $\boldsymbol{\sigma}_H(\mathbf{n})=\begin{pmatrix} n_yT_{\mathrm{Xy}} \\ -n_xT_{\mathrm{Xy}} \\ 0 \end{pmatrix}+\begin{pmatrix} 3T_{\mathrm{Xxxy}}n_y^3-3T_{\mathrm{Xxxz}}n_zn_y^2+3T_{\mathrm{Xxxy}}n_x^2n_y+3T_{\mathrm{Xyzz}}n_z^2n_y+3T_{\mathrm{Xxxz}}n_x^2n_z \\ -3T_{\mathrm{Xxxy}}n_x^3-3T_{\mathrm{Xxxy}}n_y^2n_x-6T_{\mathrm{Xxxz}}n_yn_zn_x-3T_{\mathrm{Xyzz}}n_z^2n_x \\ T_{\mathrm{Zxxx}}n_x^3-3T_{\mathrm{Zxxx}}n_xn_y^2 \end{pmatrix}$ | 1st order: 1<br>3rd order: 4 | Rashba |
| 31m | $\boldsymbol{\sigma}_H(\mathbf{n})=\begin{pmatrix} n_yT_{\mathrm{Xy}} \\ -n_xT_{\mathrm{Xy}} \\ 0 \end{pmatrix}+\begin{pmatrix} 3T_{\mathrm{Xxxy}}n_y^3+3T_{\mathrm{Xxxy}}n_x^2n_y+6T_{\mathrm{Xxyz}}n_xn_zn_y+3T_{\mathrm{Xyzz}}n_z^2n_y \\ -3T_{\mathrm{Xxxy}}n_x^3+3T_{\mathrm{Xxyz}}n_zn_x^2-3T_{\mathrm{Xxxy}}n_y^2n_x-3T_{\mathrm{Xyzz}}n_z^2n_x-3T_{\mathrm{Xxyz}}n_y^2n_z \\ 3T_{\mathrm{Zxxy}}n_x^2n_y-T_{\mathrm{Zxxy}}n_y^3 \end{pmatrix}$ | 1st order: 1<br>3rd order: 4 | Rashba |
| $\bar{6}$ | $\boldsymbol{\sigma}_H(\mathbf{n})=\begin{pmatrix} 0 \\ 0 \\ 0 \end{pmatrix}+\begin{pmatrix} 3T_{\mathrm{Xxxz}}n_zn_x^2+6T_{\mathrm{Xxyz}}n_yn_zn_x-3T_{\mathrm{Xxxz}}n_y^2n_z \\ 3T_{\mathrm{Xxyz}}n_zn_x^2-6T_{\mathrm{Xxxz}}n_yn_zn_x-3T_{\mathrm{Xxyz}}n_y^2n_z \\ T_{\mathrm{Zxxx}}n_x^3+3T_{\mathrm{Zxxy}}n_yn_x^2-3T_{\mathrm{Zxxx}}n_y^2n_x-T_{\mathrm{Zxxy}}n_y^3 \end{pmatrix}$ | 1st order: 0<br>3rd order: 4 | Pure cubic texture |
| 6mm | $\boldsymbol{\sigma}_H(\mathbf{n})=\begin{pmatrix} T_{\mathrm{Xy}}n_y \\ -T_{\mathrm{Xy}}n_x \\ 0 \end{pmatrix}+\begin{pmatrix} 3T_{\mathrm{Xxxy}}n_x^2n_y+3T_{\mathrm{Xxxy}}n_y^3+3T_{\mathrm{Xyzz}}n_yn_z^2 \\ -3T_{\mathrm{Xxxy}}n_xn_y^2-3T_{\mathrm{Xxxy}}n_x^3-3T_{\mathrm{Xyzz}}n_xn_z^2 \\ 0 \end{pmatrix}$ | 1st order: 1<br>3rd order: 2 | Rashba |
| $\bar{6}$m2 | $\boldsymbol{\sigma}_H(\mathbf{n})=\begin{pmatrix} 0 \\ 0 \\ 0 \end{pmatrix}+\begin{pmatrix} 3T_{\mathrm{Xxxz}}n_x^2n_z-3T_{\mathrm{Xxxz}}n_y^2n_z \\ -6T_{\mathrm{Xxxz}}n_xn_yn_z \\ T_{\mathrm{Zxxx}}n_x^3-3T_{\mathrm{Zxxx}}n_xn_y^2 \end{pmatrix}$ | 1st order: 0<br>3rd order: 2 | Pure cubic texture |
| $\bar{6}$2m | $\boldsymbol{\sigma}_H(\mathbf{n})=\begin{pmatrix} 0 \\ 0 \\ 0 \end{pmatrix}+\begin{pmatrix} 6T_{\mathrm{Xxyz}}n_xn_yn_z \\ 3T_{\mathrm{Xxyz}}n_x^2n_z-3T_{\mathrm{Xxyz}}n_y^2n_z \\ 3T_{\mathrm{Zxxy}}n_x^2n_y-T_{\mathrm{Zxxy}}n_y^3 \end{pmatrix}$ | 1st order: 0<br>3rd order: 2 | Pure cubic texture |
| $\bar{4}$3m | $\boldsymbol{\sigma}_H(\mathbf{n})=\begin{pmatrix} 0 \\ 0 \\ 0 \end{pmatrix}+\begin{pmatrix} 3T_{\mathrm{Xxyy}}n_xn_y^2-3T_{\mathrm{Xxyy}}n_xn_z^2 \\ 3T_{\mathrm{Xxyy}}n_yn_z^2-3T_{\mathrm{Xxyy}}n_x^2n_y \\ 3T_{\mathrm{Xxyy}}n_x^2n_z-3T_{\mathrm{Xxyy}}n_y^2n_z \end{pmatrix}$ | 1st order: 0<br>3rd order: 1 | Pure cubic texture |

We also identified the groups that frequently occur in altermagnetic materials [8-11]: $P2_1/n$ (14), C2/c (15), Cmcm (63), Pnma (62), $P4_2/mnm$ (136), I4/mcm (140), R$\bar{3}$c (167), and $P6_3/mmc$ (194), all of which fall into the aforementioned category. Besides, with the help of Table S3, we can quickly identify the AHE texture for real altermagnetic materials, because it simplifies and streamlines the symmetry analysis. For example, if $\{R|\mathbf{t}\}$ operation exchanges magnetic atoms, we define its determinant as negative, and treat $R$ as improper point operations $IR$ ($I$ is the inversion symmetry) for the axial tensors $\mathbf{T}^{(2)}$ and $\mathbf{T}^{(4)}$. Conversely, if the magnetic atoms are not exchanged under $\{R|\mathbf{t}\}$, we set det($R$)=1, allowing $R$ to act as the proper point operations on $\mathbf{T}^{(2)}$ and $\mathbf{T}^{(4)}$. From there, we can identify the equivalent point groups and corresponding $\boldsymbol{\sigma}_H(\mathbf{n})$ relations in Table S3. Specific examples are provided in Sec. VI.

## Supplementary Note 5. Projection to real sphere harmonics and multipoles

Analogous to the Berry curvature multipoles in moment space [12-14] and the magnetic multipoles in real space [15-17], Eq. (12) can be interpreted as the AHE multipoles within the Néel vector space. Subsequently, we proceed to project $\boldsymbol{\sigma}_H$ into the basis of Néel vector multipoles.

The linear term of $\boldsymbol{\sigma}_H$ on $\mathbf{n}$ has three independent elements in Eq. (1), namely $n_x$, $n_y$, $n_z$. Besides, the cubic term of $\boldsymbol{\sigma}_H$ has the commutative symmetry as stated in Eq. (11), resulting in 10 independent Taylor expansion terms, namely, $n_x^3$, $n_y^3$, $n_z^3$, $n_xn_yn_z$, $n_x^2n_z$, $n_y^2 n_z$, $n_x n_z^2$, $n_y n_z^2$, $n_x n_y^2$ and $n_x^2 n_y$. Additionally, **n** is a unit vector (|**n**|=1), i.e., $n_x^2 + n_y^2 + n_z^2 = 1$. This constraint reduces the independent cubic terms to 7, which is significant as they coincide with the *f* spherical harmonics with angular momentum *l*=3. Consequently, it becomes feasible to project the anomalous Hall vector $\boldsymbol{\sigma}_H$ onto the basis of real spherical harmonics with *p* and *f* orbits. The three real *p* spherical harmonics for *l* =1 are:

$$\begin{cases} p_z = \sqrt{\dfrac{3}{4\pi}}\cos\theta, \\ p_x = \sqrt{\dfrac{3}{4\pi}}\sin\theta\cos\varphi, \\ p_y = \sqrt{\dfrac{3}{4\pi}}\sin\theta\sin\varphi, \end{cases} \tag{18}$$

where $\theta$ is the polar angle, and $\varphi$ is the azimuthal angle. The seven real *f* spherical harmonics with *l* =3 are:

$$\begin{cases} f_{z^3} = Y_3^0 = \dfrac{\sqrt{7}}{4\sqrt{\pi}}(5\cos^3\theta - 3\cos\theta), \\ f_{xz^2} = \dfrac{1}{\sqrt{2}}\left(Y_3^{-1} - Y_3^1\right) = \dfrac{\sqrt{21}}{4\sqrt{2\pi}}(5\cos^2\theta - 1)\sin\theta\cos\varphi, \\ f_{yz^2} = \dfrac{i}{\sqrt{2}}\left(Y_3^{-1} + Y_3^1\right) = \dfrac{\sqrt{21}}{4\sqrt{2\pi}}(5\cos^2\theta - 1)\sin\theta\sin\varphi, \\ f_{z(x^2-y^2)} = \dfrac{1}{\sqrt{2}}\left(Y_3^2 + Y_3^{-2}\right) = \dfrac{\sqrt{105}}{4\sqrt{\pi}}\sin^2\theta\cos\theta\cos 2\varphi, \\ f_{xyz} = \dfrac{i}{\sqrt{2}}\left(Y_3^{-2} - Y_3^2\right) = \dfrac{\sqrt{105}}{4\sqrt{\pi}}\sin^2\theta\cos\theta\sin 2\varphi, \\ f_{x(x^2-3y^2)} = \dfrac{1}{\sqrt{2}}\left(Y_3^{-3} - Y_3^{-3}\right) = \dfrac{\sqrt{35}}{4\sqrt{2\pi}}\sin^3\theta\left(\cos^2\varphi - 3\sin^2\varphi\right)\cos\varphi, \\ f_{y(3x^2-y^2)} = \dfrac{i}{\sqrt{2}}\left(Y_3^{-3} + Y_3^3\right) = \dfrac{\sqrt{35}}{4\sqrt{2\pi}}\sin^3\theta\left(3\cos^2\varphi - \sin^2\varphi\right)\sin\varphi. \end{cases} \tag{19}$$

The multipoles with *p* and *f* configurations are shown in Fig. S6. For an arbitrary real spherical harmonic $\phi_i$ in Eq. (18) and Eq. (19), it obeys:

$$\int_0^{2\pi} \mathrm{d}\varphi \int_0^{\pi} \phi_i \phi_j \sin\theta \mathrm{d}\theta = \delta_{ij}, \tag{20}$$

where $\delta_{ij}$ is the Kronecker delta function. The real spherical harmonics are orthogonal and normalized. Thus, $\{\phi_i\}$ = $\{p_z$, $p_x$, $p_y$, $f_{z^3}$, $f_{xz^2}$, $f_{yz^2}$, $f_{z(x^2-y^2)}$, $f_{xyz}$, $f_{x(x^2-3y^2)}$, $f_{y(3x^2-y^2)}\}$ forms a complete and independent basis set in the multipole space.

We utilize the relations $n_x = \sin\theta\cos\varphi$, $n_y = \sin\theta\sin\varphi$, and $n_z = \cos\theta$ to convert $\boldsymbol{\sigma}_H$ from Cartesian coordinates in Eq. (12) to spherical coordinates. The expanded coefficient of $\boldsymbol{\sigma}_H$ within the basis $\phi_j$ is:

$$\zeta_{ij} = \int_0^{2\pi} \mathrm{d}\varphi \int_0^{\pi} \sigma_i \phi_j \sin\theta \mathrm{d}\theta. \tag{21}$$

The coefficients of the $p$ sphere harmonics function represent the dipoles in Néel vector space, and the $f$ sphere harmonics function to the octupoles. Finally, the anomalous Hall vector $\boldsymbol{\sigma}_H$ can be expressed as the multipoles as follows:

$$\sigma_{Hi} = \sum_j \zeta_{ij} \phi_j. \tag{22}$$

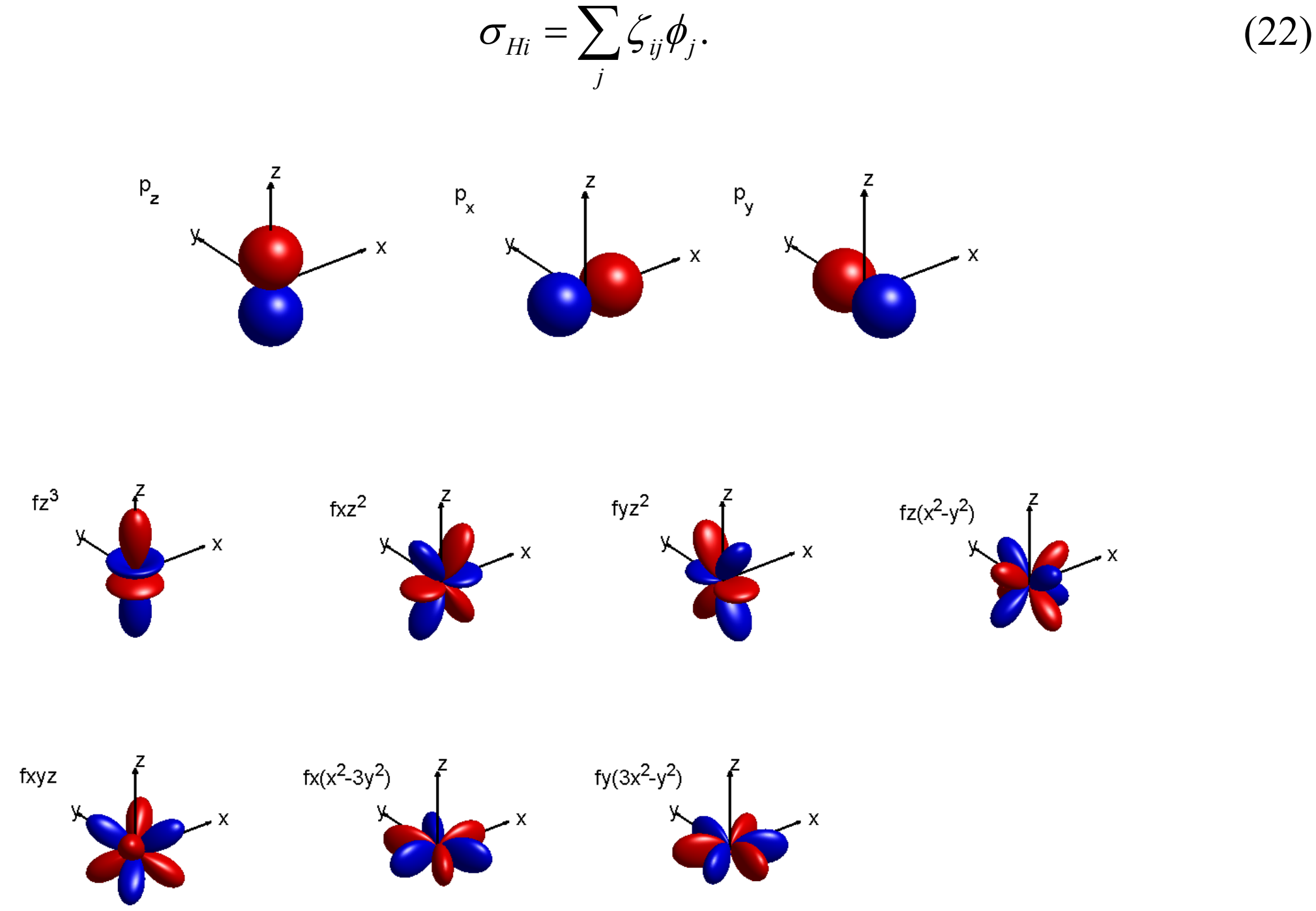

Fig. S6. Real spherical harmonics of $p$ and $f$ orbits.

The multipoles of $\boldsymbol{\sigma}_H$ in the Néel vector space reflect the complexity of the spin-orbit coupling (SOC) effect [3]. For example, the dipole $\boldsymbol{\sigma}_H$ with magnetism $\mathbf{m}$ in ferromagnet is at least first order in the spin-conversing part and second order in the spin-flipping part of the SOC effect. Besides, the $\boldsymbol{\sigma}_H$ with octupole structure demonstrates a scaling relationship with $\lambda^3$ (SOC strength) for the spin-conserving part of SOC terms, and with $\lambda^4$ for the spin-flip SOC components [3]. The verification of these conclusions needs further study.

## Supplementary Note 6. Effective TB models and calculation results

### Supplementary Note 6.1. The methods to construct the TB models and to calculate AHE conductivity

We use the *s*-orbitals of the magnetic atoms to construct the effective tight-binding (TB) Hamiltonians. First, we utilize the MagneticTB [18] program to obtain symmetry-constrained effective Hamiltonians with spin-orbit coupling (SOC) effect but without magnetism. Under the symmetry operation $\{R|\boldsymbol{t}\}$, the TB Hamiltonian satisfies:

$$P\left(\{R\,|\,\mathbf{t}\}\right)^{-1} H(\mathbf{k}) P\left(\{R\,|\,\mathbf{t}\}\right) = H^*\left(-R^{-1}\mathbf{k}\right), \tag{23}$$

where the $P(\{R|\boldsymbol{t}\})$ is the symmetry operation matrix under the orbital bases. Finally, we add the on-site magnetic exchange interactions, as follows:

$$H_{mag} = (-1)^i \sum_i \mathbf{m}\cdot\boldsymbol{\sigma} c_i^+ c_i, \tag{24}$$

where $i$ denotes the site index of magnetic atoms, $\boldsymbol{\sigma}$ represents the Pauli matrices, and $\mathbf{m} = m(\sin\theta\cos\varphi\,, \sin\theta\sin\varphi\,, \cos\theta)$ is the magnetic moments. Eq. (24) ensures that the magnetic moments of the two sublattices of altermagnets are anti-parallel all the way, i.e., there is no net magnetic moment when the Néel vector changes.

According to Kubo-Greenwood theory, we compute the coefficients for the AHE and the magneto-optical effects as follows:

$$\sigma_{Hk}(\omega) = \frac{\mathrm{i}e^2\hbar}{N_k\Omega_c}\varepsilon_{ijk}\sum_{\mathrm{k}}\sum_{n,m}\frac{f_{\mathrm{mk}}-f_{\mathrm{nk}}}{\varepsilon_{\mathrm{mk}}-\varepsilon_{\mathrm{nk}}}\frac{\langle\psi_{n\mathbf{k}}|v_i|\psi_{m\mathbf{k}}\rangle\langle\psi_{m\mathbf{k}}|v_j|\psi_{n\mathbf{k}}\rangle}{\varepsilon_{m\mathbf{k}}-\varepsilon_{n\mathbf{k}}-(\hbar\omega+i\eta)}, \tag{25}$$

where $\omega$ is the frequency of light, $\varepsilon_{ijk}$ is the Levi-Civita symbol. The anomalous Hall conductivity can be obtained when $\omega = 0$. Within the effective TB Hamiltonians, we use our own WRFP program to calculate the AHE and magnetic-optical coefficients.

### Supplementary Note 6.2. Rutile-type altermagnetic materials

A rutile-type altermagnetic material has two magnetic atoms in A (0, 0, 0) and B (1/2, 1/2, 1/2), respectively. The effective TB Hamiltonian without magnetism under the basis of $\{A\uparrow,\ A\downarrow,\ B\uparrow,\ B\downarrow\}$ is:

$$H = \begin{bmatrix} 0 & 0 & H_{13} & H_{14} \\ 0 & 0 & -H_{14}^* & H_{13} \\ \dagger & \dagger & 0 & 0 \\ \dagger & \dagger & 0 & 0 \end{bmatrix}, \tag{26}$$

where

$$H_{13} = t\sum_{ij} c_i^+ c_j = 8t\cos\left(\frac{k_x}{2}\right)\cos\left(\frac{k_y}{2}\right)\cos\left(\frac{k_z}{2}\right), \tag{27}$$

is the hopping of two magnetic atoms, and

$$H_{14} = (4+4\mathrm{i})\lambda\left(\sin\left[\frac{k_x - k_y}{2}\right] + \mathrm{i}\sin\left[\frac{k_x + k_y}{2}\right]\right)\sin\left(\frac{k_z}{2}\right), \tag{28}$$

is the SOC coupling term. This effective model is consistent with Refs. [19,20]. The band structures without the SOC effect are shown in Fig. S7(a). Whereafter, we incorporate the antiferromagnetic exchange interaction term. The magnetism induces the band splitting, as shown in Fig. S7(b). The flat band at the X-S-Y arises from the fact that TB model in Eq. (26) is too simplistic, leading to $H_{13} = 0$ and $H_{14} = 0$ at these paths. The AHE conductivities of the main text are adopted at -0.61 eV.

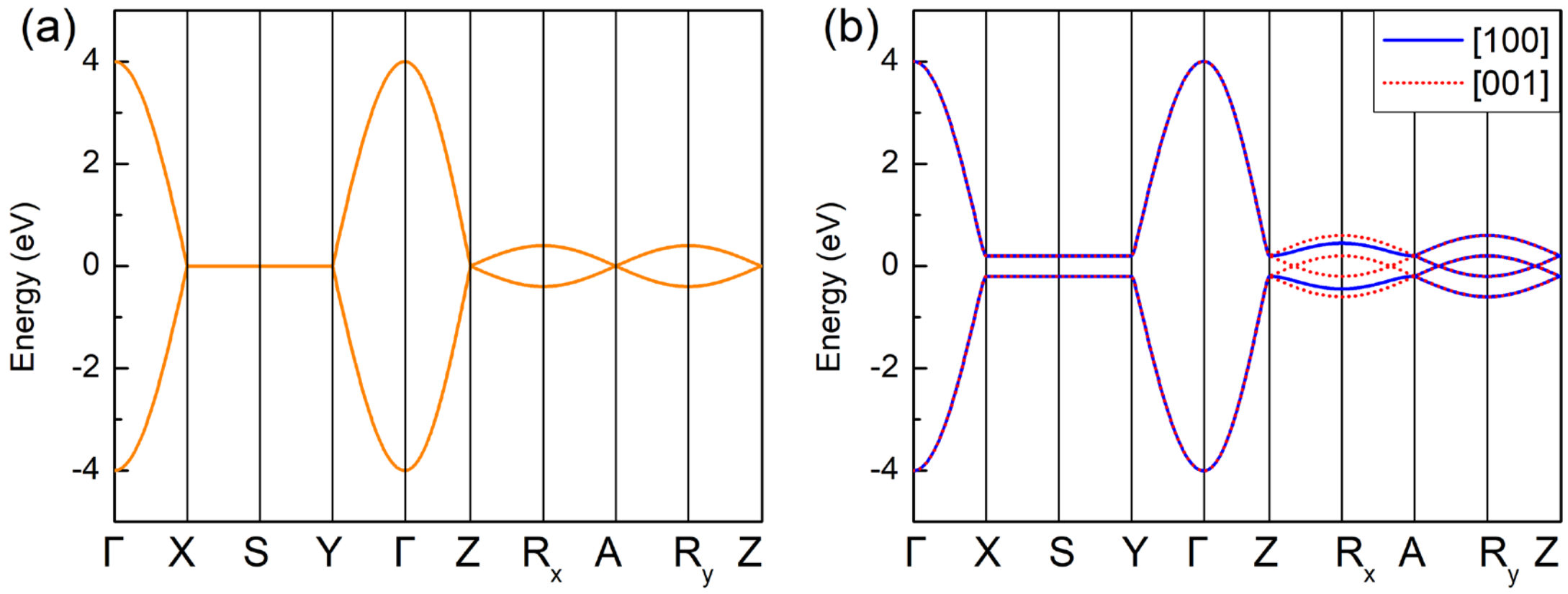

Fig. S7. Bands structures of effective TB model in Eq. (26) (a) without and (b) with magnetization. $t$=0.5 eV, $\lambda$= 0.05 eV and $m$=0.2 eV. The vectors in (b) denote the Néel vectors.

### Supplementary Note 6.3. Calcite-type altermagnetic materials

The primitive cell of calcite-type altermagnets with $R\bar{3}c$ space group is

$$\begin{cases} \mathbf{a}_1 = \frac{a}{2}\mathbf{i} + \frac{\sqrt{3}a}{6}\mathbf{j} + \frac{c}{3}\mathbf{k}, \\ \mathbf{a}_2 = -\frac{a}{2}\mathbf{i} + \frac{\sqrt{3}a}{6}\mathbf{j} + \frac{c}{3}\mathbf{k}, \\ \mathbf{a}_3 = -\frac{\sqrt{3}}{3}a\mathbf{j} + \frac{c}{3}\mathbf{k}, \end{cases} \tag{29}$$

where $a$ and $c$ are the lattice constants in the hexagonal cell. There are two magnetic atoms in each primitive cell, occupied at A (0, 0, 0) and B (1/2, 1/2, 1/2), respectively. The effective TB Hamiltonian without magnetism including the second-nearest neighbor terms (A-B, A-A) is:

$$H=\begin{bmatrix} 0 & 0 & H_{13} & H_{14} \\ 0 & 0 & -H_{14}^{*} & H_{13}^{*} \\ \dagger & \dagger & 0 & 0 \\ \dagger & \dagger & 0 & 0 \end{bmatrix}, \tag{30}$$

where

$$\begin{aligned} H_{13} &= 2(r_3+ir_1)\left[2\cos\left(\frac{k_x}{2}\right)\cos\left(\frac{k_y-k_z}{2}\right)+\cos\left(\frac{1}{2}(k_x-k_y-k_z)\right)\right] \\ &+2(t_2+it_1)\cos\left(\frac{1}{2}(k_x+k_y+k_z)\right), \end{aligned} \tag{31}$$

and

$$H_{14}=8r_2\cos\left[\frac{k_z}{2}\right]\sin\left[\frac{k_x}{2}\right]\sin\left[\frac{k_y}{2}\right]+4ir_2\left(\sqrt{3}\sin\left[\frac{k_x-k_y}{2}\right]+i\sin\left[\frac{k_x+k_y}{2}\right]\right)\sin\left[\frac{k_z}{2}\right]. \tag{32}$$

The band structures are shown in Fig. S8. The AHE conductivities of the main text are adopted at -2.86 eV.

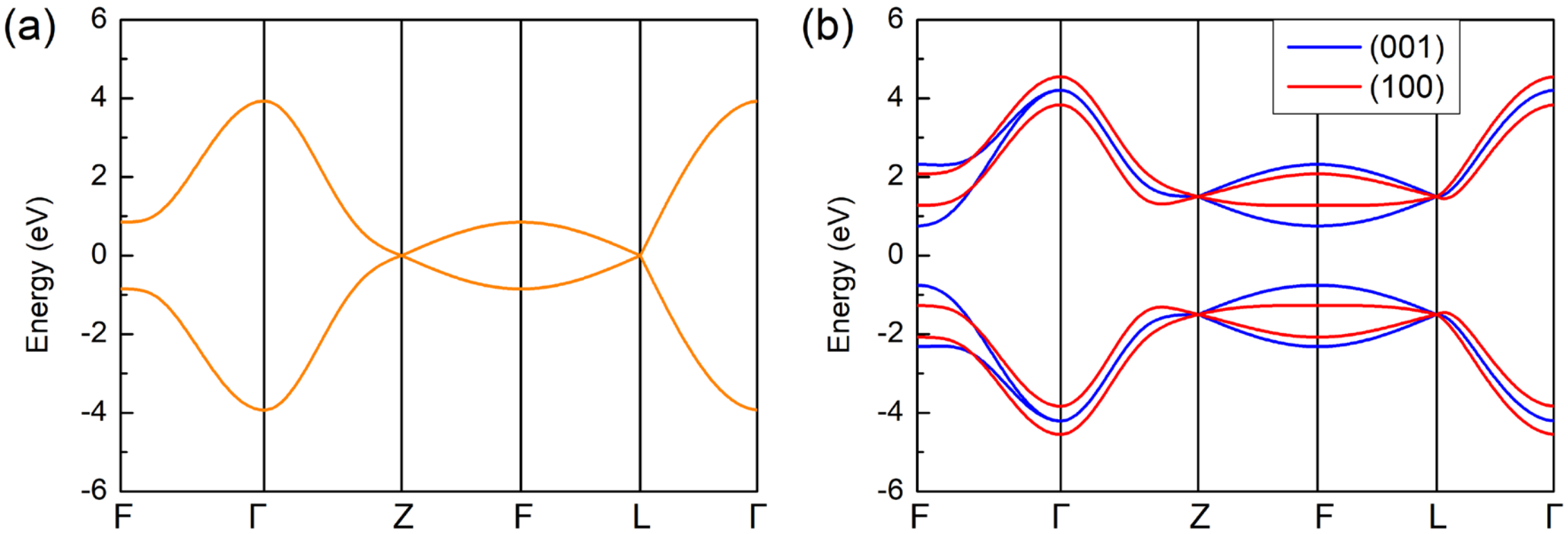

Fig. S8. Band structure of effective TB model in Eq. (30) (a) without and (b) with magnetization. $t_1$ = 0.05 eV, $t_2$ = 0.4 eV, $r_1$= 0.15 eV, $r_2$ =0.1 eV, $r_3$ = 0.5 eV and $m$=1.5 eV. The vectors in (b) denote the Néel vectors.

## Supplementary Note 6.4. NiAs-type altermagnetic materials

There are two magnetic atoms in each unit cell for the NiAs-type altermagnetic material, located at A (0, 0, 0) and B (0, 0, 1/2), respectively. The TB Hamiltonian without magnetism is:

$$H=\begin{bmatrix} H_{11} & 0 & H_{13} & H_{14} \\ 0 & H_{11} & -H_{14}^{*} & H_{13} \\ \dagger & \dagger & H_{11} & 0 \\ \dagger & \dagger & 0 & H_{11} \end{bmatrix}, \tag{33}$$

where

$$H_{11}=2t_2\left[\cos\left(k_x\right)+\cos\left(k_y\right)+\cos\left(k_x+k_y\right)\right] \tag{34}$$

is the hopping of six in-plane next-nearest neighbors, and

$$\begin{aligned} H_{13}&=2\left[t_1+2t_3\left(\cos\left(k_x\right)+\cos\left(k_y\right)+\cos\left(k_x+k_y\right)\right)\right]\cos\left(\frac{k_z}{2}\right) \\ &-4i\lambda_1\left[\sin\left(k_x\right)+\sin\left(k_y\right)-\sin\left(k_x+k_y\right)\right]\sin\left(\frac{k_z}{2}\right), \end{aligned} \tag{35}$$

where $t_1$ is the nearest neighbor hopping, $t_3$ is the third-nearest neighbor hopping, and $\lambda_1$ is the SOC strength. $H_{14}$ represents the SOC effect between different sublattices, and its expression is:

$$H_{14}=\frac{4}{3}\lambda_2\left[-2i\sqrt{3}\cos\left(k_x\right)+i\left(3\mathrm{i}+\sqrt{3}\right)\cos\left(k_y\right)+\left(3+i\sqrt{3}\right)\cos\left(k_x+k_y\right)\right]\cos\left(\frac{k_z}{2}\right). \tag{36}$$

Finally, we incorporate the antiferromagnetic exchange interaction term using Eq. (24). The calculated energy bands are shown in Fig. S9. The anomalous Hall conductivity of the main text is adopted at -3 eV.

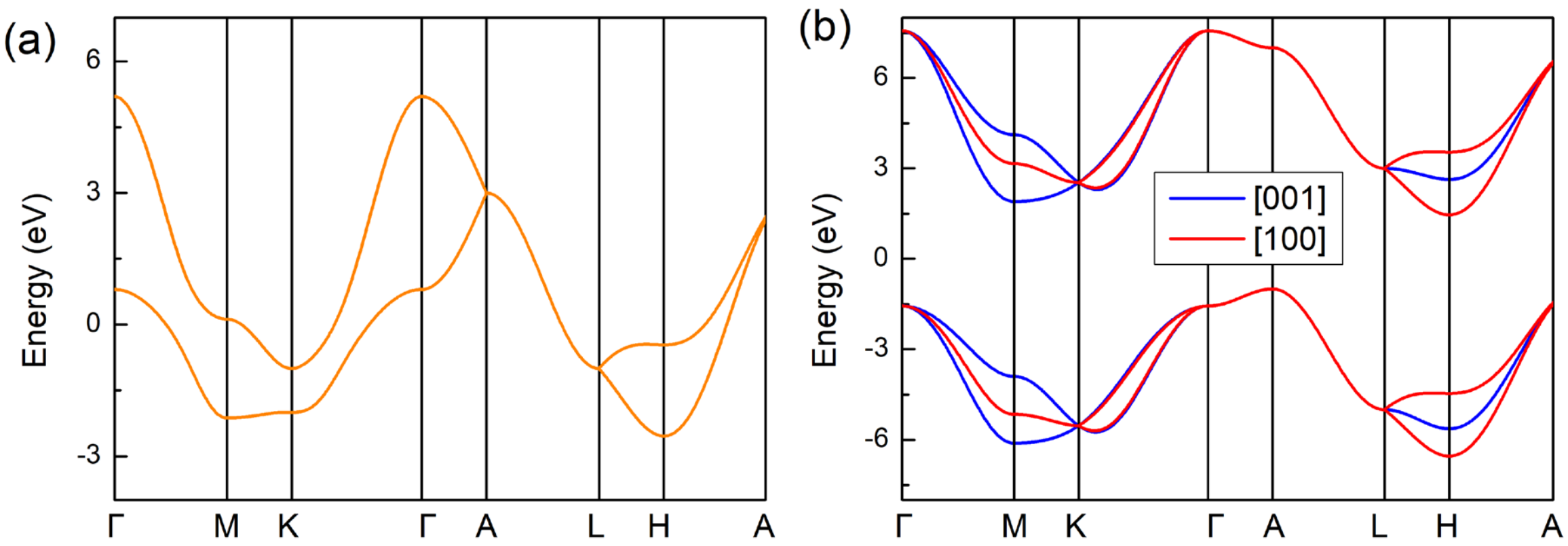

Fig. S9. Band structures of NiAs-type crystal with effective TB model in Eq. (33) (a) without and (b) with magnetization. $t_1$=0.2 eV, $t_2$ =0.5 eV, $t_3$ =0.15 eV, $\lambda_1$=-0.1 eV, $\lambda_2$=0.12 eV, $m$=4.0 eV. The vectors in (b) denote the Néel vectors.

### Supplementary Note 6.5. Perovskite-type altermagnetic materials

The effective TB Hamiltonian without magnetism is given by:

$$H=\begin{pmatrix} 0 & 0 & H_{13} & H_{14} & H_{15} & H_{16} & H_{17} & H_{18} \\ 0 & 0 & -H_{14}^{*} & H_{13}^{*} & -H_{16}^{*} & H_{15}^{*} & -H_{18}^{*} & H_{17}^{*} \\ \dagger & \dagger & 0 & 0 & H_{17} & -H_{18}^{*} & H_{15}^{*} & -H_{16}^{*} \\ \dagger & \dagger & 0 & 0 & 0 & 0 & H_{16} & H_{15} \\ \dagger & \dagger & \dagger & \dagger & 0 & 0 & H_{57} & H_{58} \\ \dagger & \dagger & \dagger & \dagger & \dagger & \dagger & -H_{58}^{*} & H_{57}^{*} \\ \dagger & \dagger & \dagger & \dagger & \dagger & \dagger & 0 & 0 \\ \dagger & \dagger & \dagger & \dagger & \dagger & \dagger & 0 & 0 \end{pmatrix}, \tag{37}$$

where

$$H_{13}=4\cos\left(\frac{k_x}{2}\right)\left[\left(-is_2+s_6\right)\cos\left(\frac{k_y-k_z}{2}\right)+\left(is_4+s_8\right)\cos\left(\frac{k_y+k_z}{2}\right)\right], \tag{38}$$

where $s_6$ and $s_8$ are the hopping parameters, while $s_2$ and $s_4$ are the SOC coefficients. $H_{14}$ represents the spin-orbit coupling between the (1/2, 0, 0) and (0, 1/2, 1/2) atoms:

$$H_{14}=2\begin{pmatrix} \left(is_3+s_7\right)\cos\left[\frac{1}{2}\left(k_x-k_y-k_z\right)\right]+\left(-is_1+s_5\right)\cos\left[\frac{1}{2}\left(k_x+k_y-k_z\right)\right] \\ +\left(is_1+s_5\right)\cos\left[\frac{1}{2}\left(k_x-k_y+k_z\right)\right]+\left(-is_3+s_7\right)\cos\left[\frac{1}{2}\left(k_x+k_y+k_z\right)\right] \end{pmatrix}. \tag{39}$$

The hopping between $\left(\frac{1}{2},0,0\right)$ and $(\frac{1}{2},\ \frac{1}{2},\ 0)$ is:

$$H_{15}=2\left(it_2+t_3\right)\cos\left(\frac{k_y}{2}\right), \tag{40}$$

$$H_{16}=-2it_1\cos\left(\frac{k_y}{2}\right). \tag{41}$$

The hopping between $(\frac{1}{2},0,\ 0)$ and $(0,\ 0,\ \frac{1}{2})$ is:

$$H_{17}=4r_4\cos\left(\frac{k_x}{2}\right)\cos\left(\frac{k_z}{2}\right)+4ir_2\sin\left(\frac{k_x}{2}\right)\sin\left(\frac{k_z}{2}\right), \tag{42}$$

$$H_{18}=4i\left(r_1+ir_3\right)\cos\left(\frac{k_x}{2}\right)\cos\left(\frac{k_z}{2}\right), \tag{43}$$

where $r_4$ is the hopping coefficient, $r_1$, $r_2$ and $r_3$ is the SOC coefficients.

The hopping terms between $(\frac{1}{2},\frac{1}{2},0)$ and $\left(0,0,\ \frac{1}{2}\right)$ atoms are

$$H_{57}=4\cos\left(\frac{k_x}{2}\right)\left[\left(\mathrm{i}s_2+s_6\right)\cos\left(\frac{k_y+k_z}{2}\right)+\left(-\mathrm{i}s_4+s_8\right)\cos\left(\frac{k_y-k_z}{2}\right)\right], \tag{44}$$

$$H_{58}=2\begin{pmatrix}(is_3+s_7)\cos\left[\frac{1}{2}(k_x-k_y+k_z)\right]+(-is_1+s_5)\cos\left[\frac{1}{2}(k_x+k_y+k_z)\right]\\ +(is_1+s_5)\cos\left[\frac{1}{2}(k_x-k_y-k_z)\right]+(-is_3+s_7)\cos\left[\frac{1}{2}(k_x+k_y-k_z)\right]\end{pmatrix}. \quad (45)$$

Here, we use $t_3$=0.55 eV, $r_4$=0.54 eV, $s_6$=0.53 eV, $s_8$=0.52 eV, $t_1$=0.0497 eV, $t_2$=0.0499 eV, $r_1$=0.0501 eV, $r_2$=0.502 eV, $r_3$=0.503 eV, $s_1$=0.0501 eV, $s_2$=0.0503 eV, $s_3$=0.0505 eV, $s_4$=0.0507 eV, $s_5$=0.0509 eV, $s_7$=0.0511 eV, $m$=5.0 eV in Eq. (37).

## Supplementary Note 7. Anomalous-Hall Néel textures of typical altermagnets

### Supplementary Note 7.1. Rutile-type altermagnetic materials

Rutile-type altermagnetic materials, which belong to the space group P4$_2$/mnm (No. 136), such as $RuO_2$ [19,21-23], $NiF_2$, $CoF_2$, and $MnF_2$ [24]. In these AFM materials, the two magnetic atoms occupy the 2a sites, specifically at (0, 0, 0) and (1/2, 1/2, 1/2), as illustrated in Fig. S10(a). These two magnetic atoms exhibit a site symmetry of mmm, meaning they are linked by the screw $4_2$, $m_z$, $g_x$, and $g_y$ symmetries rather than the inversion symmetry.

The parent space group of rutile-type altermagnets is P4$_2$/mnm (No. 136). This space group encompasses 16 fundamental operations, and $4_2$, $g_x$, $g_y$, and $m_z$ serve as the generation operations. The transformation of anomalous Hall vector $\boldsymbol{\sigma}_H$ and Néel vector $\mathbf{n}$ under these operations is given in Table S4, and the equivalent point operations are also listed. According to equivalent point operations, the constraints on $\mathbf{T}^{(2)}$ and $\mathbf{T}^{(4)}$ are equivalent to those of points group $\bar{4}$m2.

Table S4. Transformations of $\boldsymbol{\sigma}_H$ and $\mathbf{n}$ of rutile-type altermagnets with the space group of P4$_2$/mnm. $\sigma_i$ ($i$=$x$, $y$ or $z$) is the anomalous Hall conductivity along the $i$ direction.

| $\{R\|\mathbf{t}\}$ | ($\sigma_{Hx}$, $\sigma_{Hy}$, $\sigma_{Hz}$) | Switch opposite magnetic atoms or not? | ($n_x$, $n_y$, $n_z$) | Equivalent point operation |
|---|---|---|---|---|
| $g_x$ | $(\sigma_{Hx},-\sigma_{Hy},-\sigma_{Hz})$ | yes | $(-n_x,n_y,n_z)$ | $m_x$ |
| $g_y$ | $(-\sigma_{Hx},\sigma_{Hy},-\sigma_{Hz})$ | yes | $(n_x,-n_y,n_z)$ | $m_y$ |
| $m_z$ | $(-\sigma_{Hx},-\sigma_{Hy},\sigma_{Hz})$ | no | $(-n_x,-n_y,n_z)$ | $2_z$ |
| $4_2$ | $(\sigma_{Hy},-\sigma_{Hx},\sigma_{Hz})$ | yes | $(-n_y,n_x,-n_z)$ | $\bar{4}$ |

Based on Table S4 and the AHE-texture code, we can determine $\boldsymbol{\sigma}_H$ within first-order and third-order terms of $\mathbf{n}$ as follows:

$$
\begin{aligned}
\boldsymbol{\sigma}_H(\mathbf{n}) &= (An_y, An_x, 0) + (an_x^2 n_y + bn_y^3 + cn_y n_z^2, bn_x^3 + an_x n_y^2 + cn_x n_z^2, dn_x n_y n_z) \\
&= \left(\alpha p_y - \beta f_{yz^2} + \gamma f_{y(3x^2\text{-}y^2)}, \alpha p_x - \beta f_{xz^2} - \gamma f_{x(3x^2\text{-}y^2)}, \lambda f_{xyz}\right),
\end{aligned} \tag{46}
$$

where $\alpha = \frac{2}{5}\sqrt{\frac{\pi}{3}}(5A + a + 3b + c)$ is the AHE dipole coefficient in the Néel vector space, $\beta = \frac{1}{5}\sqrt{\frac{2\pi}{21}}(a + 3b - 4c)$, $\gamma = \sqrt{\frac{2\pi}{35}}(a - b)$ and $\lambda = 2d\sqrt{\frac{\pi}{105}}$ are the AHE octupole coefficients in Néel vector space. This result is constant with that in Table S3.

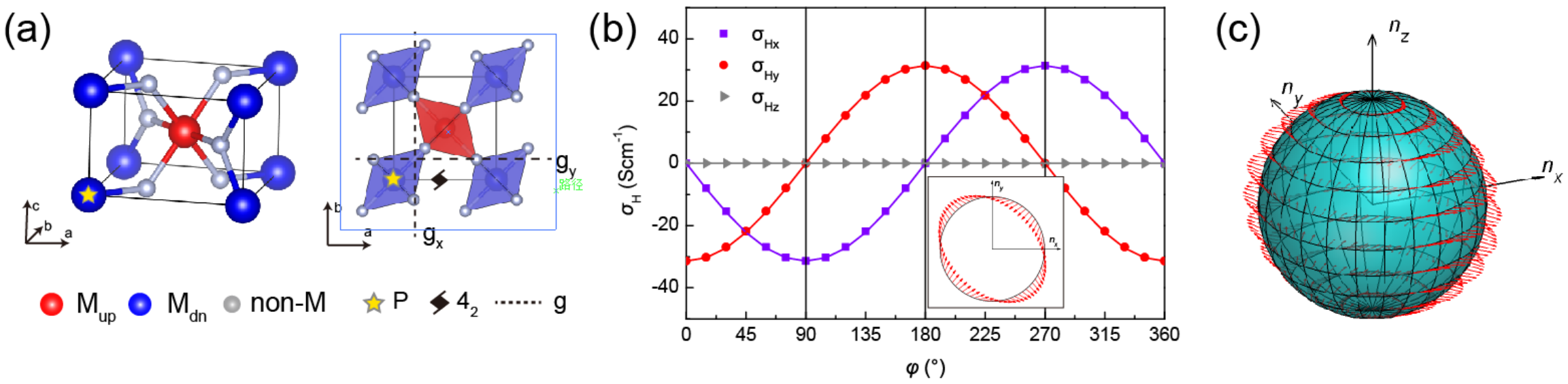

Fig. S10. (a) Crystal structure of rutile-type altermagnetic materials (left: side view, right: top view). (b) Angle dependence of $\boldsymbol{\sigma}_H$ on $\mathbf{n}$ in the *xoy* plane. (c) AHNT of rutile-type altermagnets. The inset in (b) is the Dresselhaus-like AHNT in the xoy plane. The points in (b) are the numerical data calculated by effective TB model, and the curves are the analytical $\boldsymbol{\sigma}_H(\mathbf{n})$ function fitted by Eq. (46).

The calculated $\boldsymbol{\sigma}_H(\mathbf{n})$ based on the TB model is shown in Fig. S10 (b), where the first-order terms dominate, and $\sigma_x$ and $\sigma_y$ oscillate in a sinusoidal and cosine manner for azimuth angle $\varphi$ in the *xoy* plane. It is clear that our calculation is consistent with Eq. (46), and also aligns with the previous first-principles calculations of $RuO_2$ [25], where the specific expression of $\boldsymbol{\sigma}_H(\mathbf{n})$ and AHNT are absent in that work. We can outline the $\boldsymbol{\sigma}_H$ in the Néel vector space $\mathbf{n}$, as illustrated in the inset of Fig. S10 (c), which closely resembles the Dresselhaus spin texture in the momentum space.

Except for the rutile-type altermagnets, the Dresselhaus AHNTs can also emerge in the tetragonal $ZrMn_2Ge_4O_{12}$ with the P4/nbm (No. 125) space group and $LiFe_2F_6$ with $P4_2$/mnm (No. 136) space group, as listed in Table I.

### Supplementary Note 7.2. Calcite-type altermagnets

Calcite-type alter magnets have the $R\overline{3}c$ (No. 167) space group symmetry. The transformation of $\boldsymbol{\sigma}_H$ and $\mathbf{n}$ under symmetry operations is given in Table S5, and the equivalent point operations are also listed. According to equivalent point operations, the constraints on $\mathbf{T}^{(2)}$ and $\mathbf{T}^{(4)}$ are equivalent to those of point group 3m1.

Based on the extrinsic parameter method, the symmetry-constrained $\boldsymbol{\sigma}_H$ is:

$$\boldsymbol{\sigma}_H(\mathbf{n}) = \begin{pmatrix} T_{Xy} n_y \\ -T_{Xy} n_x \\ 0 \end{pmatrix} + \begin{pmatrix} 3T_{Xxxy}\left(n_x^2+n_y^2\right)n_y + 3T_{Xxyz}\left(n_x^2-n_y^2\right)n_z + 3T_{Xyzz}n_y n_z^2 \\ -3T_{Xxxy}\left(n_x^2+n_y^2\right)n_x - 6T_{Xxyz}n_x n_y n_z - 3T_{Xyzz}n_x n_z^2 \\ T_{Zxxy}n_x(n_x^2-3n_y^2) \end{pmatrix}, \tag{47}$$

where $T_{Xy}$, $T_{Xxxy}$, $T_{Xxxz}$, $T_{Xyzz}$, $T_{Zxxx}$ are undetermined coefficients. This result is consistent with that in Table S3.

Eq. (47) can be rewritten as the following multipoles:

$$\sigma_{Hx} = \frac{2}{5}\sqrt{\frac{\pi}{3}}\left(5T_{Xy}+12T_{Xxxy}+3T_{Xyzz}\right)p_y + 4\sqrt{\frac{3\pi}{35}}T_{Xxxz}f_{z\left(x^2\text{-}y^2\right)} - \frac{4}{5}\sqrt{\frac{6\pi}{7}}\left(T_{Xxxy}-T_{Xyzz}\right)f_{yz^2}, \tag{48}$$

$$\sigma_{Hy} = -\frac{2}{5}\sqrt{\frac{\pi}{3}}\left(5T_{Xy}+12T_{Xxxy}+3T_{Xyzz}\right)p_x - 4\sqrt{\frac{3\pi}{35}}T_{Xxxz}f_{xyz} + \frac{4}{5}\sqrt{\frac{6\pi}{7}}\left(T_{Xxxy}-T_{Xyzz}\right)f_{xz^2}, \tag{49}$$

$$\sigma_{Hz} = 4\sqrt{\frac{2\pi}{35}}T_{Zxxx}f_{x\left(x^2\text{-}3y^2\right)}. \tag{50}$$

Since the coefficients of the first-order terms of $\sigma_{Hx}$ and $\sigma_{Hy}$ are opposite, this results in a Rashba-like AHNTs, which is shown in Fig. 3 of the main text.

Table S5. Transformations of $\boldsymbol{\sigma}_H$ and **n** of calcite-type altermagnets with space group $R\bar{3}c$ (No. 167).

| $\{R\|\mathbf{t}\}$ | ($\sigma_{Hx}$, $\sigma_{Hy}$, $\sigma_{Hz}$) | Switch opposite magnetic atoms or not? | ($n_x$, $n_y$, $n_z$) | Equivalent point operation |
|---|---|---|---|---|
| $g_x$ | $(\sigma_{Hx}, -\sigma_{Hy}, -\sigma_{Hz})$ | yes | $(-n_x, n_y, n_z)$ | $m_x$ |
| 3 | $\left(-\frac{\sigma_{Hx}}{2}+\frac{\sqrt{3}}{2}\sigma_{Hy}, -\frac{\sqrt{3}}{2}\sigma_{Hx}-\frac{1}{2}\sigma_{Hy}, \sigma_{Hz}\right)$ | no | $\left(-\frac{n_x}{2}+\frac{\sqrt{3}}{2}n_y, -\frac{\sqrt{3}}{2}n_x-\frac{1}{2}n_y, n_z\right)$ | 3 |
| $\bar{1}$ | ($\sigma_{Hx}$, $\sigma_{Hy}$, $\sigma_{Hz}$) | no | ($n_x$, $n_y$, $n_z$) | 1 |

### Supplementary Note 7.3. NiAs-type altermagnetic materials

Hexagonal NiAs-type altermagnetic materials include $\alpha$-MnTe [26,27], CrSb [28,29], etc.. In these materials, two opposite-spin magnetic atoms occupy 2a sites at (0, 0, 0) and (0, 0, 1/2). This kind of altermagnetic material exhibits an A-type AFM ground state, where the spins align ferromagnetically within the *ab* plane and antiferromagnetically between adjacent layers, as shown in Fig. S11(a).

The parent space group of NiAs-type altermagnets is $P6_3/mmc$ (No. 194). It has 24 basic operations, and $6_3$, $m_x$, $g_y$, and $m_z$ are the generation operations. The transformation manners of the anomalous Hall vector $\boldsymbol{\sigma}_H$ and the Néel vector **n** under these operations are outlined in Table S6, and the equivalent point operations are also listed. According to equivalent point operations, the constraints on $\mathbf{T}^{(2)}$ and $\mathbf{T}^{(4)}$ are equivalent to those of points group $\overline{6}2$m.

Table S6. Transformations of $\boldsymbol{\sigma}_H$ and $\mathbf{n}$ in NiAs-type altermagnets with space group P6$_3$/mmc.

| $\{R\|\mathbf{t}\}$ | ($\sigma_{Hx}$, $\sigma_{Hy}$, $\sigma_{Hz}$) | Switch opposite magnetic atoms or not? | ($n_x$, $n_y$, $n_z$) | Equivalent point operation |
|---|---|---|---|---|
| $m_x$ | $(\sigma_{Hx}, -\sigma_{Hy}, -\sigma_{Hz})$ | no | $(n_x, -n_y, -n_z)$ | $2_x$ |
| $g_y$ | $(-\sigma_{Hx}, \sigma_{Hy}, -\sigma_{Hz})$ | yes | $(n_x, -n_y, n_z)$ | $m_y$ |
| $m_z$ | $(-\sigma_{Hx}, -\sigma_{Hy}, \sigma_{Hz})$ | yes | $(n_x, n_y, -n_z)$ | $m_z$ |
| $6_3$ | $\left(\frac{\sigma_{Hx}}{2}+\frac{\sqrt{3}}{2}\sigma_{Hy}, -\frac{\sqrt{3}}{2}\sigma_{Hx}+\frac{1}{2}\sigma_{Hy}, \sigma_{Hz}\right)$ | yes | $\left(-\frac{n_x}{2}-\frac{\sqrt{3}}{2}n_y, \frac{\sqrt{3}}{2}n_x-\frac{1}{2}n_y, -n_z\right)$ | $\overline{6}$ |

Based on the above characters, we can drive $\boldsymbol{\sigma}_H$ within Néel vector $\mathbf{n}$ polynomials up to third-order terms with the help of AHE-texture code. $\boldsymbol{\sigma}_H$ can be written as:

$$\begin{aligned}\boldsymbol{\sigma}_H(\mathbf{n}) &= \left[2an_xn_yn_z, a\left(n_x^2-n_y^2\right)n_z, c(3n_x^2n_y-n_y^3)\right] \\ &= \left[a\sqrt{\frac{\pi}{105}}f_{xyz}, a\sqrt{\frac{\pi}{105}}f_{(x^2-y^2)z}, c\sqrt{\frac{2\pi}{35}}f_{(3x^2-y^2)y}\right],\end{aligned} \tag{51}$$

where the linear order terms are forbidden by symmetry. The cubic terms correspond to the octupoles in the Néel parameter spaces. This result is consistent with that in Table S3.

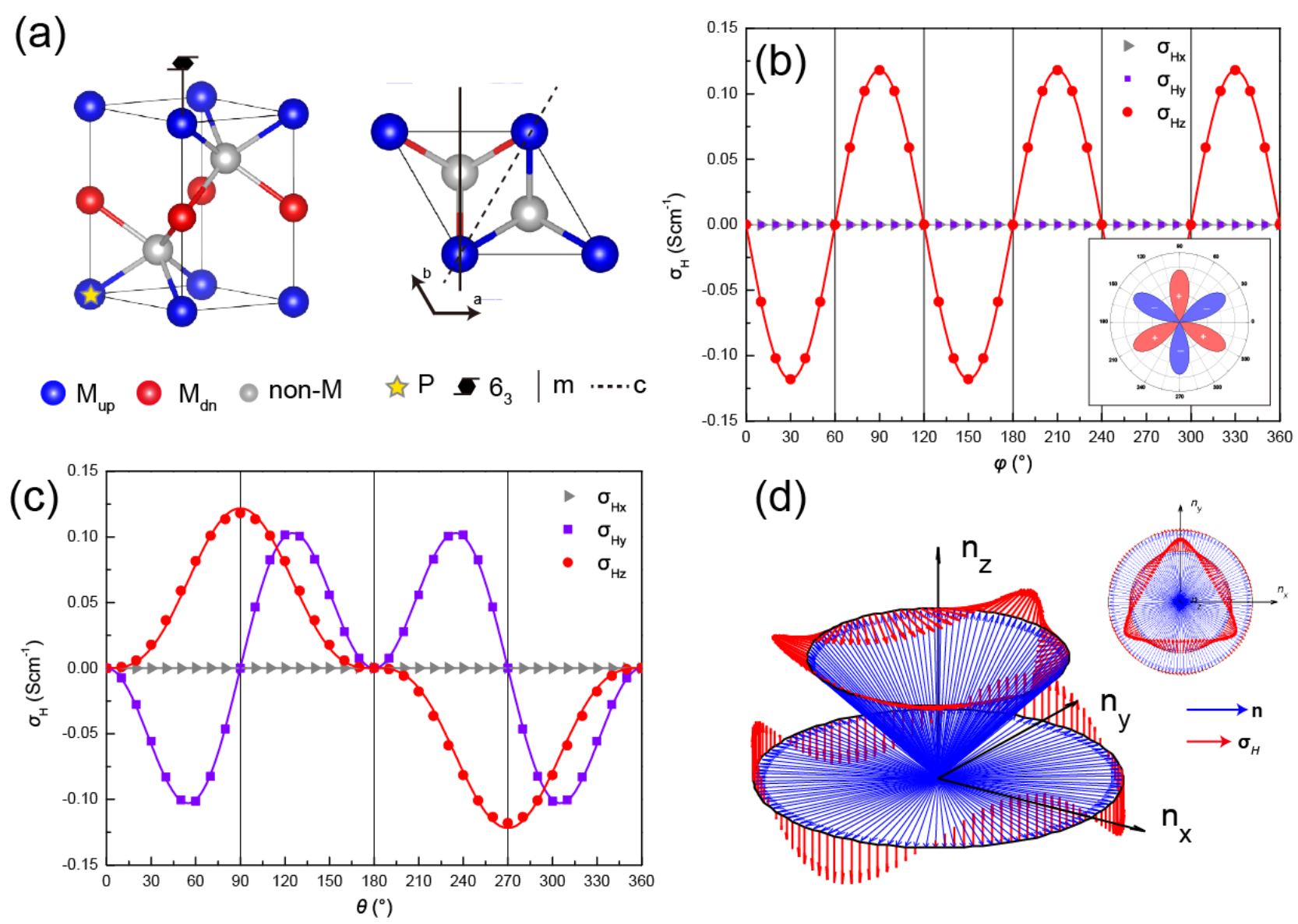

Fig. S11. (a) Crystal structure of NiAs-type altermagnets (left: side view, right: top view). Angle dependence of $\boldsymbol{\sigma}_H$ in the (b) *xoy* and (c) *yoz* plane. Inset in (b) is the magnitude of $\sigma_z$ with $\varphi$. The calculated data are fitted by Eq. (51) with lines. (d) AHNT in the Néel vector space. The insert in (d) shows the top view.

The $\boldsymbol{\sigma}_H(\mathbf{n})$ in the *xoy* and *yoz* planes derived from the TB model is presented in Fig. S11 (b) and (c), respectively. The $\boldsymbol{\sigma}_H$ only has a *z*-component when $\mathbf{n}$ lies in the *xoy* plane, and it displays the three-fold rotational symmetry with $\sigma_z = c\sin 3\varphi$, as depicted in insect of Fig. S11(b). Our results are in good agreement with both previous theoretical calculations [26,30] and AHE [26] and X-ray magnetic circular dichroism (XMCD) [31,32] measurements, where $\boldsymbol{\sigma}_H$ in the whole Néel vector space is not exposed therein. In the *yoz* plane (where $n_x$=0), $\boldsymbol{\sigma}_H = [0, -an_y^2 n_z, -cn_y^3]$. Therefore, $\sigma_y$ reaches a minimum at $\theta = \tan^{-1}(\sqrt{2}) \approx 54.7°$, and $\sigma_z$ is maximized at $\theta$= 90°, as shown in Fig. 4(c). This behavior well aligns with Eq. (51). Finally, we plot its AHNT in Fig. S11 (d). Note that our method reveals altermagnets, such as $Ba_3NiRu_2O_9$, and $CrNb_4S_8$, with the purely cubic AHNTs, as listed in Table I of the main text.

## Supplementary Note 7.4. Perovskite-type altermagnetic materials

Although the symmetry restrictions imposed by space group operations on the anomalous Hall vector $\boldsymbol{\sigma}_H$ are the same in the three magnetic configurations, the constraints on the Néel vector $\mathbf{n}$ differ (see Table S7). Therefore, the relations of $\boldsymbol{\sigma}_H - \mathbf{n}$ differ in different magnetic configurations. The expressions of $\boldsymbol{\sigma}_H(\mathbf{n})$ up to the third-order are:

$$\begin{cases} \boldsymbol{\sigma}_H^A(\mathbf{n}) = (An_y, Bn_x, 0) + (an_x^2 n_y + bn_y^3 + cn_y n_z^2, dn_x^3 + en_x n_y^2 + fn_x n_z^2, gn_x n_y n_z), \\ \boldsymbol{\sigma}_H^C(\mathbf{n}) = (An_z, 0, Bn_x) + (an_x^2 n_z + bn_y^2 n_z + cn_z^3, dn_x n_y n_z, en_x^3 + fn_x n_y^2 + gn_x n_z^2), \\ \boldsymbol{\sigma}_H^G(\mathbf{n}) = (0, An_z, Bn_y) + (an_x n_y n_z, bn_x^2 n_z + cn_y^2 n_z + dn_z^3, en_x^2 n_y + fn_y^3 + gn_y n_z^2). \end{cases} \tag{52}$$

Table S7. Transformations of $\boldsymbol{\sigma}_H$ and $\mathbf{n}$ in perovskite-type altermagnets with the Pnma space group. The "yes" or "no" denotes whether $\{R|\mathbf{t}\}$ switch opposite magnetic atoms or not.

| Symmetry $\{R\|\mathbf{t}\}$ | $(\sigma_{Hx}, \sigma_{Hy}, \sigma_{Hz})$ | A-type | C-type | G-type |
|---|---|---|---|---|
| $n_a$ | $(\sigma_{Hx}, -\sigma_{Hy}, -\sigma_{Hz})$ | Yes $(-n_x, n_y, n_z)$ | Yes $(-n_x, n_y, n_z)$ | No $(n_x, -n_y, -n_z)$ |
| $m_b$ | $(-\sigma_{Hx}, \sigma_{Hy}, -\sigma_{Hz})$ | Yes $(n_x, -n_y, n_z)$ | No $(-n_x, n_y, -n_z)$ | Yes $(n_x, -n_y, n_z)$ |
| $a_c$ | $(-\sigma_{Hx}, -\sigma_{Hy}, \sigma_{Hz})$ | No $(-n_x, -n_y, n_z)$ | Yes $(n_x, n_y, -n_z)$ | Yes $(n_x, n_y, -n_z)$ |

The optical Hall conductivity $\boldsymbol{\sigma}_H(\omega)$ for three kinds of magnetic configuration are given in Fig. S12.

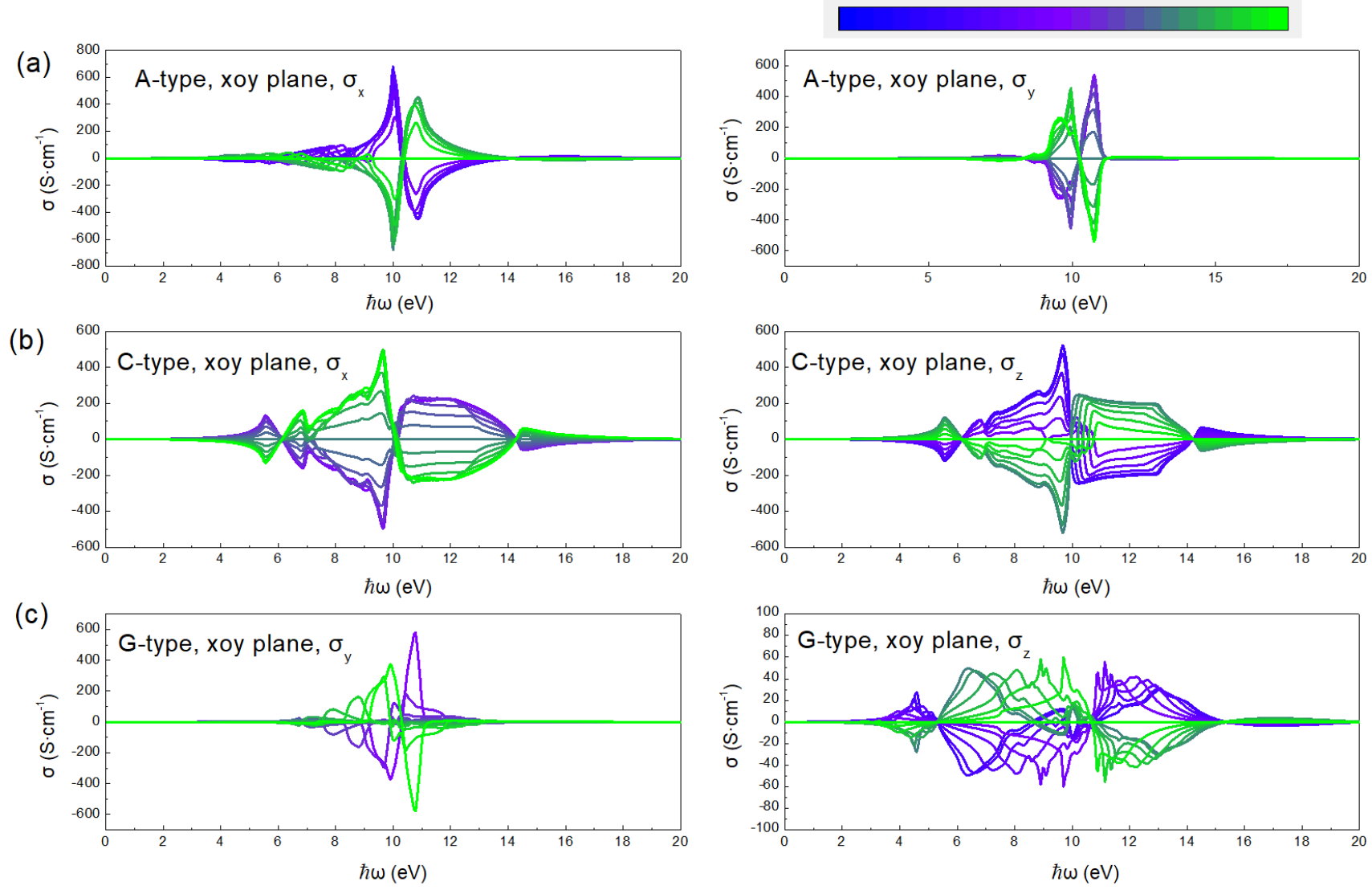

Fig. S12 The optical Hall conductivity $\boldsymbol{\sigma}_H(\omega)$ of perovskite-type altermagnetic materials $ABO_3$ with the TB model in Supplementary Note 6.5.

## Supplementary Note 8. Magnetic group method for determining the anomalous Hall vector.

Table S8. Symmetry-constrained anomalous Hall vector $\boldsymbol{\sigma}_H$ in magnetic point groups.

| Crystallographic system | MPG | anomalous Hall vector $\boldsymbol{\sigma}_H$ |
|---|---|---|
| Triclinic | $1, \bar{1}$ | $(\sigma_{Hx}, \sigma_{Hy}, \sigma_{Hz})$ |
| Monoclinic | Original point groups: $2, 2/m(2//y), m(m \perp y)$ | $(0, \sigma_{Hy}, 0)$ |
| | Black-White groups: $2', m', 2'/m'$ | $(\sigma_{Hx}, 0, \sigma_{Hz})$ |
| Orthorhombic | Black-White groups: $2'2'2, m'm'2, m'm'm$ | $(0, 0, \sigma_{Hz})$ |
| | Black-White group: $m'm2'$ | $(0, \sigma_{Hy}, 0)$ |
| Tetragonal, Trigonal, Hexagonal | Original point groups: $4, \bar{4}, 4/m, 3, \bar{3}, 6, \bar{6}, 6/m$ | $(0, 0, \sigma_{Hz})$ |
| | Black-White groups: $42'2', 4m'm', \bar{4}2'm', 4/mm'm'$ $32', 3m', \bar{3}m', 62'2', 6m'm', \bar{6}m'2', 6/mm'm'$ | |

Since anomalous Hall vector $\boldsymbol{\sigma}_H$ is a pseudovector, the symmetry constraints for $\boldsymbol{\sigma}_H$ under the operations of $R$ and $IR$ are the same:

$$IR\boldsymbol{\sigma}_H = R\boldsymbol{\sigma}_H, \quad (53)$$

where $I$ is the inversion operation. For an anti-unitary point operation $R' = TR$, $\boldsymbol{\sigma}_H$ is transformed as

$$R'\boldsymbol{\sigma}_H = -R\boldsymbol{\sigma}_H, \quad (54)$$

because time-reversal operation $T$ reverses $\boldsymbol{\sigma}_H$. Based on the above analysis, there are 31 magnetic point groups, including 13 original and 18 black-white magnetic point groups, that possess the nonzero $\boldsymbol{\sigma}_H$, which are listed in Table S8 [33].

When the magnetic moments of an altermagnets are arranged along a specific direction, we can identify its magnetic space group and magnetic point group, and subsequently determine the direction of the anomalous Hall vector based on Table S8. Table S9, Table S10, and Table S11 provide detailed judgments for rutile-type, NiAs-type, and perovskite-type altermagnets. The results align with those from MTENSOR in Bilbao Crystallographic Server (https://www.cryst.ehu.es/cgi-bin/cryst/programs/mtensor.pl). The results align with our "extrinsic parameter" method, and the latter enables the establishment of the relationships among different magnetic orientations.

Table S9. Magnetic space groups and corresponding anomalous Hall vectors $\boldsymbol{\sigma}_H$ for rutile altermagnets materials with different Néel vectors.

| **Néel vector** | **Magnetic space group** | $\boldsymbol{\sigma}_H$ |
|---|---|---|
| [100] ($x$) | Pn'nm' | $(0, \sigma_{Hy}, 0)$ |
| [010] ($y$) | Pnn'm' | $(\sigma_{Hx}, 0, 0)$ |
| [110] | Cmm'm' | $(\sigma_{Hx}, \sigma_{Hx}, 0)$ |
| [001] ($z$) | $P4_2'/mnm'$ | 0 |
| [$a$0$b$] ($xoz$) | $P2_1/c.1$ | $(0, \sigma_{Hy}, 0)$ |

Table S10. Magnetic space groups and corresponding anomalous Hall vectors $\boldsymbol{\sigma}_H$ for NiAs-type altermagnets with different Néel vectors.

| **Néel vector** | **Magnetic space group** | $\boldsymbol{\sigma}_H$ |
|---|---|---|
| [100] ($x$) | Cmcm.1 | × |
| [120] ($y$) | Cm'c'm | $(0, 0, \sigma_{Hz})$ |
| [001] ($z$) | $P6_3'/m'm'c$ | × |
| [121] | C2'/m' | $(0, \sigma_{Hy}, \sigma_{Hz})$ |

Table S11. Magnetic space groups and corresponding anomalous Hall vectors $\boldsymbol{\sigma}_H$ for perovskite-type altermagnets with different Néel vectors. The symbols in the brackets denote the nonzero anomalous Hall conductivity components.

| | A-type | C-type | G-type |
|---|---|---|---|
| [100] ($x$) | Pn'ma' ($\sigma_{Hy}$) | Pn'm'a ($\sigma_{Hz}$) | Pnma.1 (×) |
| [010] ($y$) | Pnm'a' ($\sigma_{Hx}$) | Pnma.1 (×) | Pn'm'a ($\sigma_{Hz}$) |
| [001] ($z$) | Pnma.1 (×) | Pnm'a' ($\sigma_{Hx}$) | Pn'ma' ($\sigma_{Hy}$) |

It is noteworthy that the magnetic group method cannot determine the $\boldsymbol{\sigma}_H - \mathbf{n}$ relation when the Néel vector is along a general direction. This is because that $\boldsymbol{\sigma}_H$ is the function of magnetic group in the methodology, and magnetic groups is not a good parameter. The magnetic group method encounters difficulties in addressing this issue, primarily due to treating the magnetic order parameters as intrinsic structural parameters, like crystal structures. On the contrary, the extrinsic parameter method takes the magnetic order as an extrinsic parameter rather than an intrinsic structural parameter.

## Supplementary Note 9. The application of AHNTs

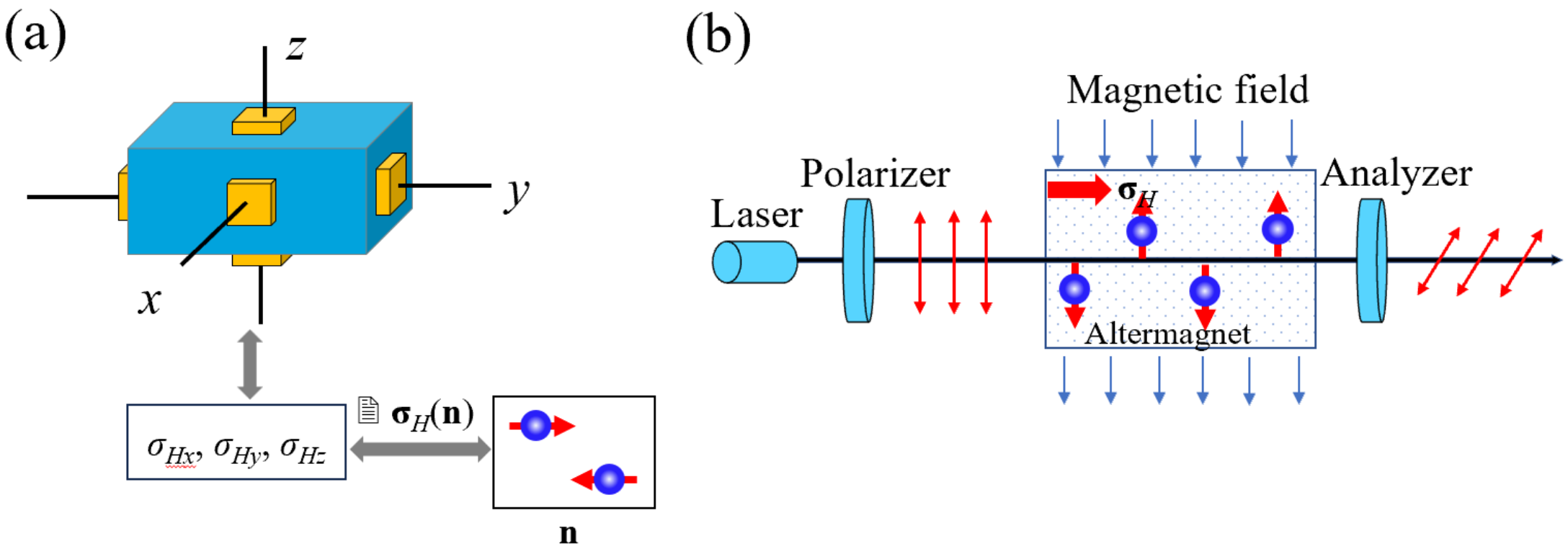

Fig. S13. (a) Schematic diagram to determine the Néel vector in an altermagnet with AHE or magneto-optical effect. $\sigma_{Hx}$, $\sigma_{Hy}$ and $\sigma_{Hz}$ are the anomalous Hall conductivity along the $x$-, $y$-, and $z$-directions, respectively. (b) A model for vertical magneto-optical isolator by utilizing the universal relationship $\boldsymbol{\sigma}_H \perp \mathbf{n}$ in altermagnets.

Based on the $\boldsymbol{\sigma}_H - \mathbf{n}$ relation, we can deduce the orientation of the Néel vector $\mathbf{n}$ more conveniently, as illustrated in Fig. S13(a), because of the analytical $\boldsymbol{\sigma}_H(\mathbf{n})$. Moreover, we can develop a vertical magneto-optical isolator, as depicted in Fig. S13(b). We can alter the $\boldsymbol{\sigma}_H$ in the horizontal direction by controlling the direction of the Néel vector in the vertical direction, thereby adjusting the magnitude of the Faraday effect and switching the on-off light. This design may offer advantages over traditional isolators, where the magnetic field is applied along the light's path, thereby limiting the size of the magnet.